\documentclass[structabstract]{aa}  
\usepackage{graphicx}
\usepackage{natbib}
\usepackage{txfonts}
\begin{document}
   \title{UV-driven chemistry in simulations of the interstellar medium}
   \subtitle{I. Post-processed chemistry with the Meudon PDR code}
   \author{F. Levrier\inst{1} \and F. Le Petit\inst{2} \and P. Hennebelle\inst{1} \and P. Lesaffre\inst{1} \and M. Gerin\inst{1} \and E. Falgarone\inst{1}
} 

\institute{LERMA/LRA - ENS Paris - UMR 8112 du CNRS, 24 rue Lhomond 75231 Paris CEDEX 05, France
\and 
LUTH - Observatoire de Paris, France
}
   \date{Received ...}
 
  \abstract
{Observations have long demonstrated the molecular diversity of the diffuse interstellar medium (ISM). Only now, with the advent of high-performance computing, does it become possible for numerical simulations of astrophysical fluids to include a treatment of chemistry, in order to faithfully reproduce the abundances of the many observed species, and especially that of CO, which is used as a proxy for molecular hydrogen. When applying photon-dominated region (PDR) codes to describe the UV-driven chemistry of uniform density cloud models, it is found that the observed abundances of CO are not well reproduced.}
   {Our main purpose is to estimate the effect of assuming uniform density on the line-of-sight in PDR chemistry models, compared to a more realistic distribution for which total gas densities may well vary by several orders of magnitude. A secondary goal of this paper is to estimate the amount of molecular hydrogen which is not properly traced by the CO ($J=1\to 0$) line, the so-called "dark molecular gas".}
   {We use results from a magnetohydrodynamical (MHD) simulation as a model for the density structures found in a turbulent diffuse ISM with no star-formation activity. The Meudon PDR code is then applied to a number of lines of sight through this model, to derive their chemical structures.}
   {It is found that, compared to the uniform density assumption, maximal chemical abundances for H$_2$, CO, CH and CN are increased by a factor $\sim 2-4$ when taking into account density fluctuations on the line of sight. The correlations between column densities of CO, CH and CN with respect to those of H$_2$ are also found to be in better overall agreement with observations. For instance, at $N(\mathrm{H}_2)\gtrsim 2~10^{20}~\mathrm{cm}^{-2}$, while observations suggest that $\mathrm{d}[\log N(\mathrm{CO})]/\mathrm{d}[\log N(\mathrm{H}_2)]\simeq 3.07\pm 0.73$, we find $\mathrm{d}[\log N(\mathrm{CO})]/\mathrm{d}[\log N(\mathrm{H}_2)]\simeq 14$ when assuming uniform density, and $\mathrm{d}[\log N(\mathrm{CO})]/\mathrm{d}[\log N(\mathrm{H}_2)] \simeq 5.2$ when including density fluctuations.} 
   {}

   \keywords{ISM : structure -- ISM : clouds -- ISM : photon-dominated region (PDR)
               }

   \maketitle

\section{Introduction}

The interstellar medium (ISM) is a complex system : its structure and dynamics are governed by the interaction of many processes covering a wide range of scales, involving both micro- and macrophysics. To understand how the ISM works is also essential on the pathway to star and planetary system formation. This field has seen much progress in recent years, both on the observational side, with the results from the Herschel Space Observatory~\citep{pilbratt10,degraauw10}, and on the theoretical side, with ever-improving numerical simulations \citep[see e.g.][]{banerjee09,vazquez10}, which now consistently treat self-gravity, thermodynamics and magnetohydrodynamics (MHD). 

The new challenge for numerical simulations of the ISM is now to incorporate some treatment of chemistry and grain physics, in order to compare with observations of atomic and molecular lines. One possibility, followed for instance by \citet{glover10} is to perform multi-fluid simulations, each fluid representing a chemical species and being coupled to the others via a network of reactions. This approach has the advantage that it naturally solves for the time-dependent distribution and dynamics of the various species in three-dimensional space, but its computational cost requires making use of simplifying assumptions, most notably a small number of species and reactions. For instance, \citet{glover10} treat a network of 218 reactions for 32 species, 13 of which are assumed to be in instantaneous chemical equilibrium, so that only 19 are actually fully treated as time-dependent quantities.

In this paper, we follow a different approach~: we post-process a single-fluid MHD simulation with the Meudon PDR code. This has the advantage of providing a full chemical network (99 species, 1362 reactions), but on the other hand, it implies a one-dimensional, steady-state treatment. With this approach, our main goal is to discuss the effects of realistic density fluctuations on PDR chemistry, since previous studies have focused on uniform density models, as in \citet{lepetit06}, or on simplified clumpy models, such as those by \citet{wolfire10}. A second goal of this study is to estimate the amount of "dark molecular gas" that can be expected from observations of the diffuse ISM, i.e. gas where hydrogen is mostly in the form of H$_2$, but where CO is too scarce to be seen in the usual tracer that is the ($J=1\to 0$) line, given current sensitivities. We wish to compare this estimate with available observations~\citep{grenier05,leroy07,abdo10,velusamy10} and models \citep{wolfire10}.

The paper is organised as follows : Section~\ref{sec:tools} describes the MHD simulation used and presents the PDR code in its current state, while section~\ref{sec:method} gives an overview of the post-processing method used to combine the two. Section~\ref{sec:extraction} discusses the lines of sight selected for running our analysis. Results and comparison to observations are presented in section~\ref{sec:full}. Section~\ref{sec:conclusions} offers further discussion and a summary. Details on using the Meudon PDR code with density profiles and our strategy for post-processing results can be found in the appendices at the end of the paper.



\section{Tools}
\label{sec:tools}
\subsection{The MHD simulation}
\label{secmhd}
As a model of the turbulent diffuse interstellar medium, we use data from a magnetohydrodynamical (MHD) simulation performed by \citet{hennebelle08} using the RAMSES code \citep{teyssier02,fromang06}. A notable advantage of this code resides in its adaptive mesh refinement (AMR) capabilities, making it able to locally reach extremely high spatial resolutions. The initial setup of the simulation is a cube of homogeneous warm neutral gas (WNM) of atomic gas (a 10:1 mixture of H and He) with density $n_H=1~\textrm{cm}^{-3}$ and temperature $T=8000~\textrm{K}$. The cube is $L=50~\mathrm{pc}$ on each side, and two converging flows of WNM are injected from opposing faces along the $X$ axis with relative velocity $\Delta V_X\sim 40~\mathrm{km.s}^{-1}$, which means that the Mach number of each flow with respect to the ambient WNM is $\mathcal{M}\sim 2$. Transverse and longitudinal velocity modulations are imposed, with amplitudes roughly equal to the mean flow velocity ($20~\mathrm{km.s}^{-1}$). Periodic boundary conditions are applied on the remaining four faces. The simulation starts on a $256^3$ grid, with two extra levels of refinement based on density thresholds, so that the effective resolution of the simulation is $\sim$ 0.05 pc. For our purposes, we regridded the data regularly on a $1024^3$ cube, therefore using the maximum resolution over the whole domain. A magnetic field is present in the simulation, and is initially parallel to the $X$ axis with an intensity of about 5~$\mu\textrm{G}$, consistent with observational values at these densities~\citep{crutcher10}. The converging flows collide near the midplane $X=0$ about $1~\textrm{Myr}$ into the simulation, and as the total mass grows from $\sim 3000~M_\odot$ initially to about 10 times this value at the end of the simulation ($t=13.11~\textrm{Myr}$), gravity eventually takes over and, combined with the effects of thermal instability \citep{field65,hennebelle00,koyama02,heitsch05,hennebelle07,banerjee09}, leads to the formation of cold dense clumps ($n_H> 100~\textrm{cm}^{-3}$, while $T\sim 10-50~\textrm{K}$) within a much more diffuse and warm interclump medium ($n_H\sim 1-10~\textrm{cm}^{-3}$ and $T\sim 10^3-10^4~\textrm{K}$). This occurs at $t\sim 12~\textrm{Myr}$. See \citet{hennebelle08} for a more detailed description. Fig.~\ref{FigColDens} shows the column density of the gas viewed along the $X$ axis, which is that of the incoming flows. 

\subsection{The Meudon PDR code}
\label{sec:PDRintro}
The Meudon PDR code ({\tt http://pdr.obspm.fr/}) is a publicly available set of routines \citep{lebourlot93,lepetit06,gonzalez08} whose purpose is to describe the UV-driven chemistry of interstellar clouds, and in particular of photon-dominated regions (PDRs). It is a steady-state one-dimensional code in which a plane-parallel slab of gas and dust is illuminated on either or both sides by the light from a star or by the standard Inter Stellar Radiation Field (ISRF), which is defined using expressions from \citet{mathis83} and \citet{black94}. Usually run with homogeneous gas densities, the code can also accept density profiles input via a text file. At each point along the line of sight, radiative transfer in the UV is treated to solve for the H/H$_2$ transition, using either the approximations by \citet{federman79}, or an exact method based on a spherical harmonics expansion of the specific intensity~\citep{goicoechea07}. A number of heating (photoelectric effect on grains, cosmic rays) and cooling (infrared and millimeter emission lines) processes contribute to the computation of thermal balance. Outputs of the code include gas properties such as temperature and ionization fraction, radiation energy density, chemical abundances and column densities, level populations and line intensities. The code is iterative and therefore requires the user to check whether convergence has been achieved.

\section{Method overview}

\label{sec:method}
Naively, one would like to run the PDR code on all lines of sight through the simulated cube to derive a three-dimensional chemical structure, as well as line-of-sight integrated observables (an emission map for the CII [158~$\mu$m] line, for instance). However, this is neither feasible nor desirable. 

Two computational reasons preclude this brute force approach. Firstly, the cost is prohibitive : just one global iteration of the PDR code for a single run typically completes in a couple hours on a GNU/Linux machine equipped with 4 dual-core 64-bit x86 processors and 64GB of memory. As a run usually converges in some 10 iterations, it takes about a day to process a single line of sight. The PDR code, however, has been ported on the EGEE grid\footnote{\tt http://www.eu-egee.org/}, which allows us to perform $\sim 100$ runs simultaneously in the same timeframe. Still, it is not reasonable to consider treating more than $\sim 10^3$ lines of sight in this work.

The second computational issue to consider is that the PDR code has trouble converging in low-density regions, of which there are many in the MHD simulation. Such are the vicinities of $X=\pm 25$~pc, where WNM gas $(n_H=1~\textrm{cm}^{-3})$ is entering the box, but low density regions are actually found everywhere throughout the cube (the volume filling factor of regions where $n_H\leqslant 20~\mathrm{cm}^{-3}$ is $f_v=0.96$). This convergence issue might be related to the fact that Ly-$\alpha$ emission, which is a major cooling process for $n_H\lesssim 5~\mathrm{cm}^{-3}$, is not included in the code, making the computation of thermal balance by the PDR code in these regions unreliable. 

Besides these computational hurdles, it should be noted that the code is one-dimensional, and treats a density profile as that of a plane-parallel slab of gas. Imagine then that a line of sight intercepts a dense clump of matter : the gas lying directly behind that clump will be shielded from incoming radiation, since the most energetic UV photons will have been absorbed by the clump. This leads to shadowing artifacts in regions which in reality may well be illuminated from other directions, as the ISM has a complex, fractal-like, and evolving structure.


In this paper, we deal with these artifacts in the following way : the physical conditions and chemical composition at every grid point $(X,Y,Z)$ considered in the analysis may be obtained by running the PDR code in $p$ directions going through that grid point, for instance along the main coordinate axes. Thus, each quantity $F(X,Y,Z)$ output by the code has $p$ possible values $f_1, f_2, \ldots, f_p$. This is the case, in particular, of the radiation energy density $E$, which has possible values $e_1, e_2, \ldots, e_p$. To combine the $p$ runs at this grid point, we select direction $p_0$ for which the radiation energy density there is maximum, $e_{p_0}=\max\{e_i\}_{1\leqslant i\leqslant p}$. This choice is discussed in section~\ref{sec:conclusions}. Of course $p_0$ is a function of $(X,Y,Z)$. Once this is done, we choose $F(X,Y,Z)=f_{p_0}$ for every quantity $F$ output by the code. This procedure takes better account of the porosity of the simulated structures to the ISRF, while ensuring element conservation at each grid point. Ideally, the more directions $p$, the better, but this obviously comes with an increased computational cost. In this paper, we choose $p=2$ as a compromise, running the PDR code in two orthogonal directions.

The analysis in this paper focuses on a small subset of a single simulation snapshot, which we discuss in the next section. After applying a density threshold $n_0=20~\textrm{cm}^{-3}$, which is the lowest value considered in the grid of models run by \citet{lepetit06} and therefore deemed sufficient to ensure convergence, we extract a number of one-dimensional density profiles from this thresholded subset. We run the PDR code on these profiles and combine results to derive the chemical structure of the clump.



\section{Lines of sight selected for PDR computations}
\label{sec:extraction}

   \begin{figure}
   \centering
   \includegraphics[width=10.5cm]{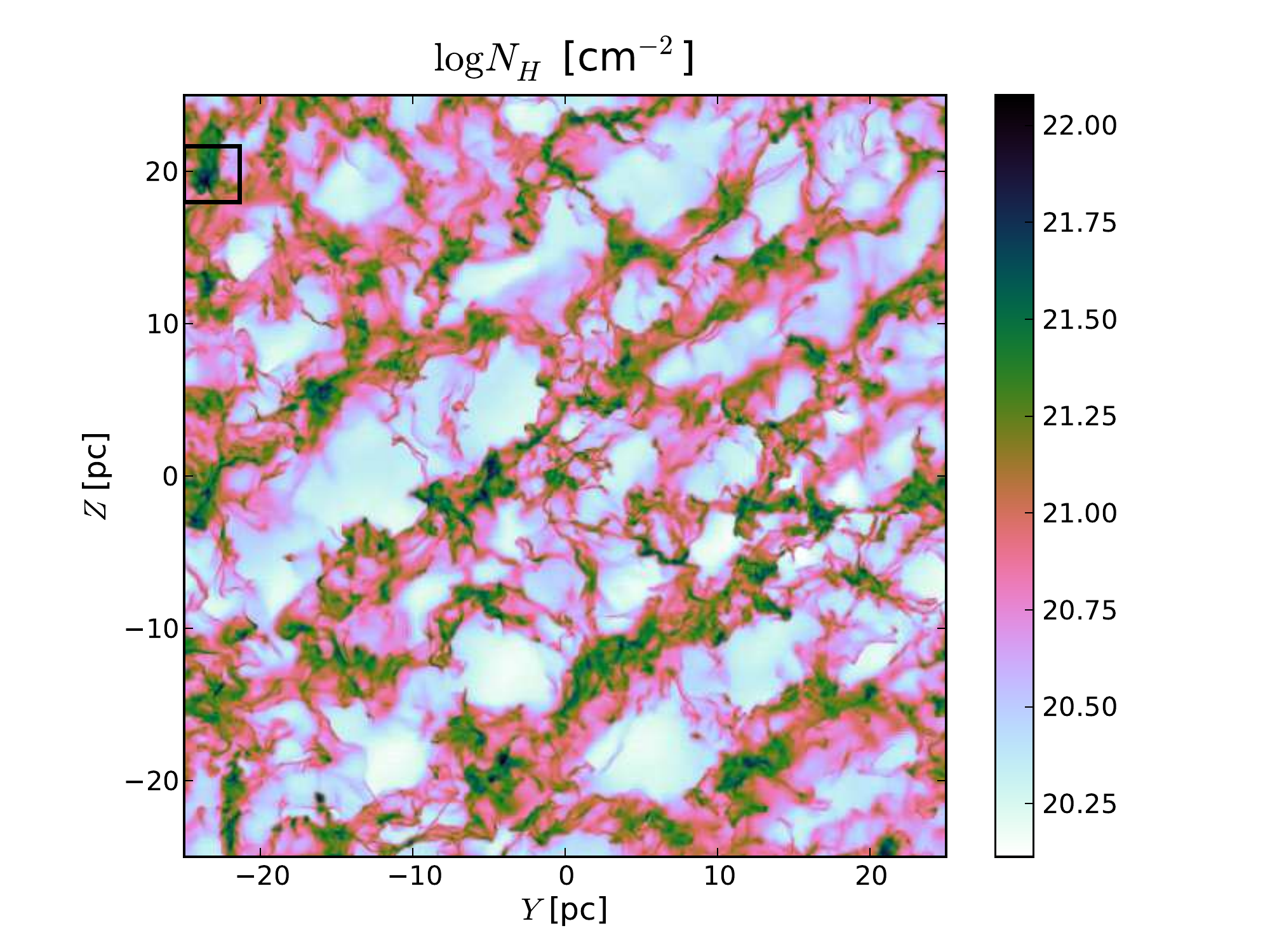}
      \caption{Total gas column density along the $X$ axis of the MHD simulation snapshot used here. The box in the upper left side marks the position of the clump selected for running our analysis.}
         \label{FigColDens}
   \end{figure}

   \begin{figure}
   \begin{center}
   \includegraphics[width=10.5cm]{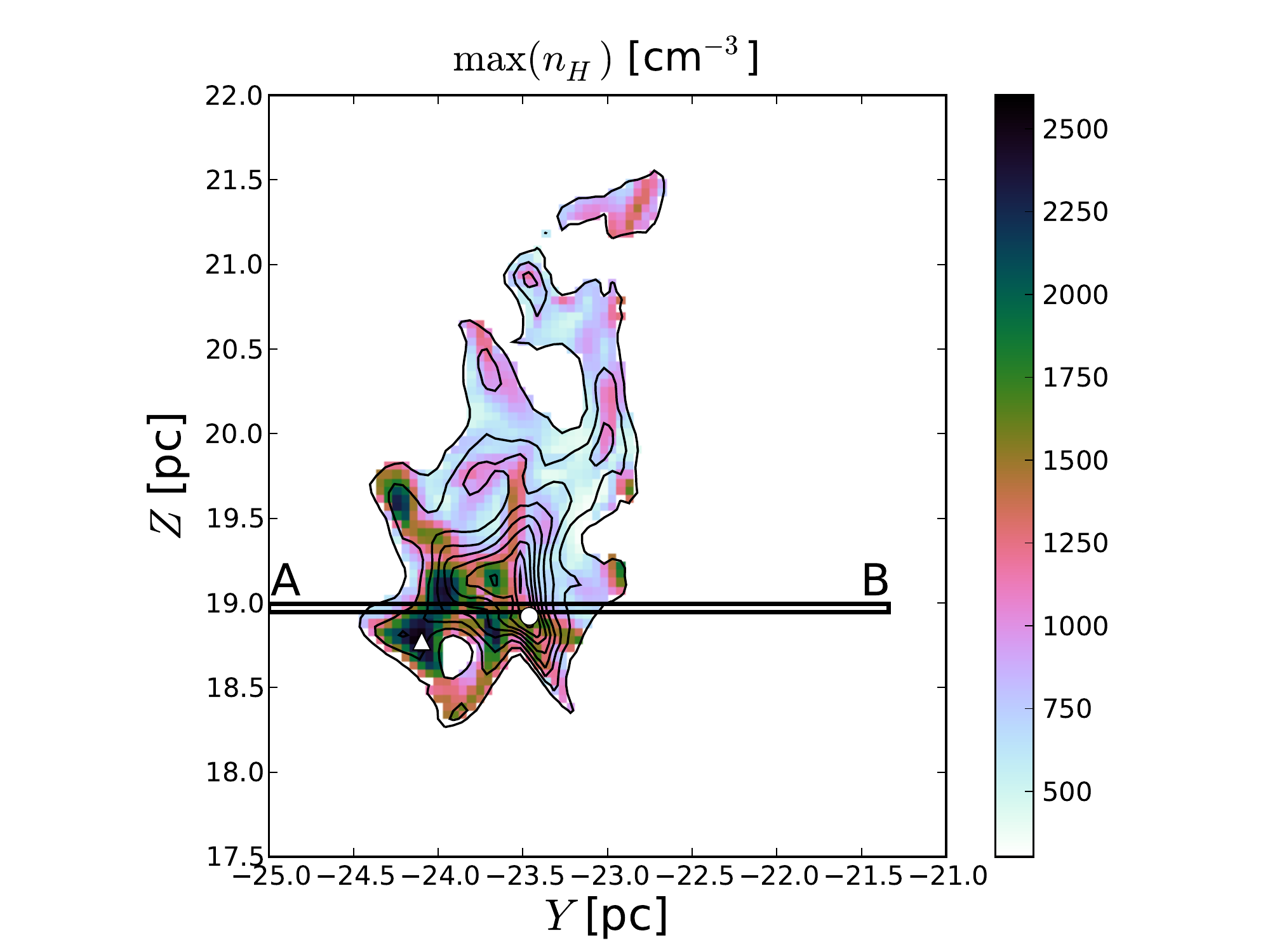}
   \end{center}
      \caption{Maximum gas density $n_H$ along the lines of sight parallel to the $X$ axis, within the selected clump. Contours show the total gas column density $N_H$ from $3~10^{21}~\mathrm{cm}^{-2}$ to $1.1~10^{22}~\mathrm{cm}^{-2}$ in steps of $10^{21}~\mathrm{cm}^{-2}$. The "clump" displayed here has a size roughly $1.5~\mathrm{pc}~\times~2.5~\mathrm{pc}$. The 2D slab of gas under study is seen projected as the single-pixel-wide {\bf AB} strip. The white triangle marks the position of the maximum value of $\max{(n_H)}$ in this region, and the white circle that of the maximum of $N_H$. They are separated by 0.65 pc.}
         \label{FigColDensClump}
   \end{figure}

\subsection{Simulation snapshot}


The snapshot chosen to run our analysis on is timed at 7.35 Myr, when the densest parts of the cloud reach $n_{\textrm{\tiny max}}\sim 9.10^3~\textrm{cm}^{-3}$. Some of the structures present in the simulation at this time are self-gravitating, but we are confident that they are still diffuse enough that the simulation snapshot is representative of a non-starforming region of the ISM. We may therefore run our analysis in the absence of any illuminating star, with the ISRF being the only source of primary UV photons. 

\subsection{Selected clump}
To select a representative subset, we may note that observational PDRs such as the Horsehead Nebula (see e.g. \citet{pety07}) are found at the edge of dense and cold clouds of gas and dust, illuminated by ambient FUV light and possibly nearby young stars. It thus makes sense to focus on a "clump", defined observationally as a connected structure with a significantly higher column density than its surroundings. We identify clumps via a friend-of-friend algorithm on the column density map along the $X$ axis (Fig.~\ref{FigColDens}), using a threshold $N_0=3.10^{21}~\mathrm{cm}^{-2}$, which corresponds to a mean density $\left<n_H\right>=n_0=20~\mathrm{cm}^{-3}$ over 50~pc.

The selected clump, which lies at the top left corner of the simulation's field, harbours an interesting feature, shown on Fig.~\ref{FigColDensClump}. That figure represents, in colour scale, the map of the maximum gas density $\max{(n_H)}$ encountered along the $X$ axis, for every line of sight within the clump. It so happens that the peak of $N_H$ does not match that of $\max{(n_H)}$, or even a local maximum of the latter. This is important to note for species which may be sensitive to the local gas density rather than to the total column density. Properties of that selected clump are listed in Table~\ref{table:2}.

\begin{table}
\caption{Properties of the selected observational clump. $\left<F\right>$ refers to the direct average and $\overline{F}$ to the density-weighted average of any quantity $F$ in this table. The mass is computed as $M=\mathcal{V}\mu m_H\left<n_H\right>$, where $\mathcal{V}$ is the 3D volume corresponding to the 2D clump, $m_H=1.66~10^{-24}~\mathrm{g}$ is the mass of the hydrogen atom and $\mu=1.4$ corresponds to a 1:10 number ratio for He with respect to H. The turbulent velocity dispersion includes all three velocity components, $\sigma_\mathrm{3D}^2=\sigma_{X}^2+\sigma_{Y}^2+\sigma_{Z}^2$.}
\label{table:2}
\centering
\begin{tabular}{l l l}
\hline\hline          
Average density  & $\left<n_H\right>$ & 30 cm$^{-3}$\\
Average temperature & $\overline{T}$ & 270 K\\
Line-of-sight centroid velocity  & $\overline{V_X}$ & -0.12 km.s$^{-1}$\\
Mass & $M$ & 124 $\mathrm{M}_\odot$\\
Turbulent dispersion & $\sigma_\mathrm{3D}$ & 1.8 km.s$^{-1}$ \\
\hline    
\end{tabular}
\end{table}


\subsection{Selected lines of sight}

\begin{figure}
   \centering
   \includegraphics[width=8.5cm]{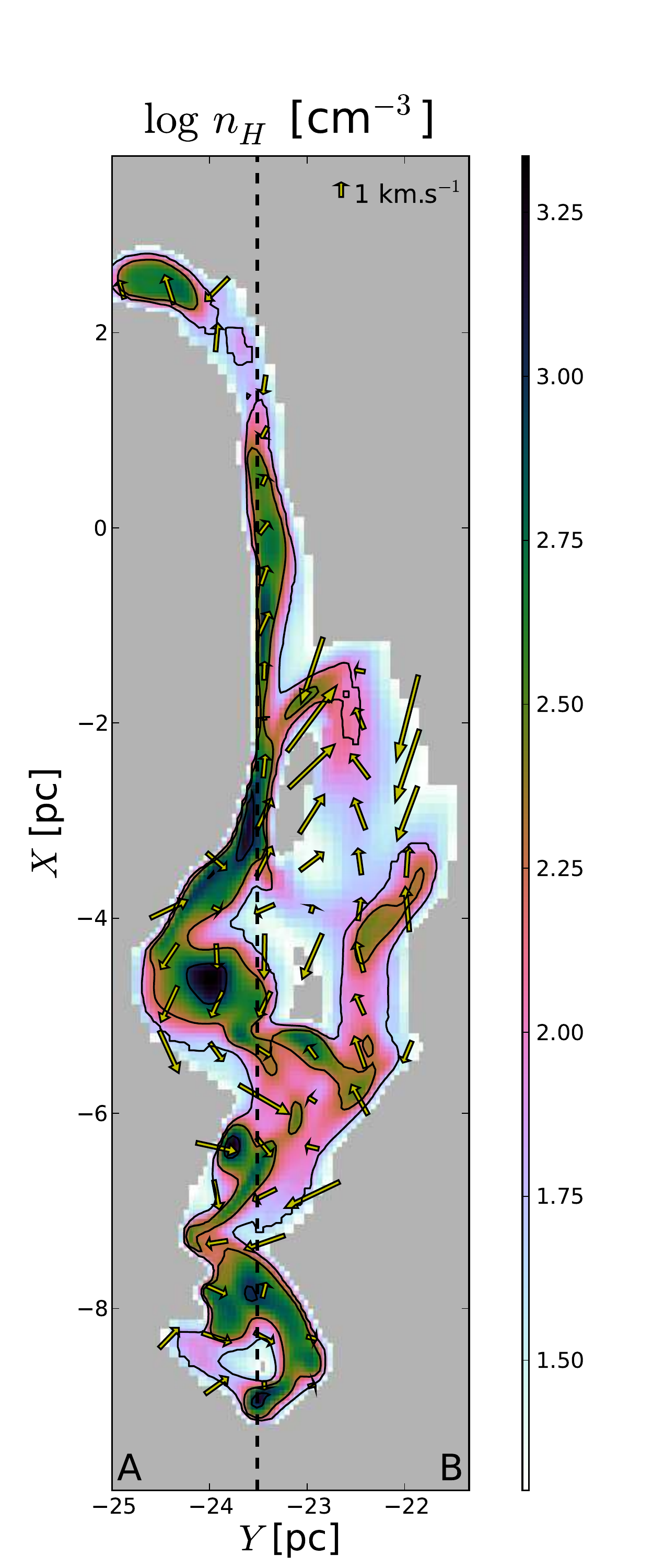}
      \caption{Structure of the gas in the 2D slab under study. This is a close-up view on the region $X\simeq 0$ where the incoming flows of WNM collide to form cold structures, and only regions where $n_H\geqslant n_0=20~\mathrm{cm}^{-3}$ are shown. The colour image shows the gas density $n_H$ in logarithmic scale, while contours show the gas temperature $T_\mathrm{MHD}$ at 20~K (higher densities), 30~K and 50~K (lower densities). The 2D projection of the velocity field $(V_X,V_Y)$ is also shown as yellow arrows, the length of which indicate the velocity modulus at that location (at the center of each arrow).  The {\bf A} and {\bf B} extremities of the observational 1D strip are indicated for reference, and the dashed line marks an example location for the extracted density profiles on which the PDR code is run (Fig.~\ref{FigSightLine}). The grey areas are outside of the domain used for PDR computations.}
         \label{Fig_D_T_Clump}
   \end{figure}

The observational clump shown on Fig.~\ref{FigColDensClump} has a size $\Delta Y\times\Delta Z\simeq 1.5~\mathrm{pc}~\times~2.5~\mathrm{pc}$. In the $X$ direction, most of its gas is located in the central $\Delta X\simeq$15~pc around $X=0$. Given the pixel size $\delta\simeq 0.05~\mathrm{pc}$, applying the PDR code on a structure of that size along the three coordinate axes requires some 25000 runs, which is beyond the scope of this work. Consequently, we restrict our study to a 2D slab of gas across the observational clump. It is projected on Fig.~\ref{FigColDensClump} as the one-pixel-wide strip {\bf AB}, which is therefore $\sim 3.7~\mathrm{pc}$ long, $\sim 0.05~\mathrm{pc}$ wide and contains 76 lines of sight along the $X$ axis. These sample a wide range of column densities, from $N_H=6.11~10^{20}~\textrm{cm}^{-2}$ (near the {\bf A} end) to $N_H=1.11~10^{22}~\textrm{cm}^{-2}$, corresponding to visual extinctions $A_V=0.33$ to $A_V=5.93$, using the conversion from total hydrogen column density $N_H=N(\mathrm{H})+2N(\mathrm{H}_2)$
\begin{equation}
A_V=\frac{R_V}{C_D}\left(\frac{N_H}{1~\mathrm{cm}^{-2}}\right)
\label{eq:1}
\end{equation}
with $R_V=3.1$ and $C_D=5.8\times 10^{21}~\mathrm{cm}^{-2}.\mathrm{mag}^{-1}$~\citep[see Table~\ref{table:1} and][]{lepetit06}. The structure of the gas on these lines of sight (Fig.~\ref{Fig_D_T_Clump}) is complex, with many small, dense and cold regions ($n_H\simeq 10^3$~cm$^{-3}$, $T\simeq 20$~K) interconnected via a filamentary structure and embedded within a much more diffuse and warm medium ($n_H\simeq 1$~cm$^{-3}$, $T\simeq 10^4$~K). The overdense regions ($n_H\geqslant n_0$) are located near the midplane of the simulation ($-9~\mathrm{pc}\lesssim X\lesssim 2~\mathrm{pc}$), where the flows collide and the gas condenses into cold structures. 

\begin{figure}
   \centering
   \includegraphics[width=9.5cm]{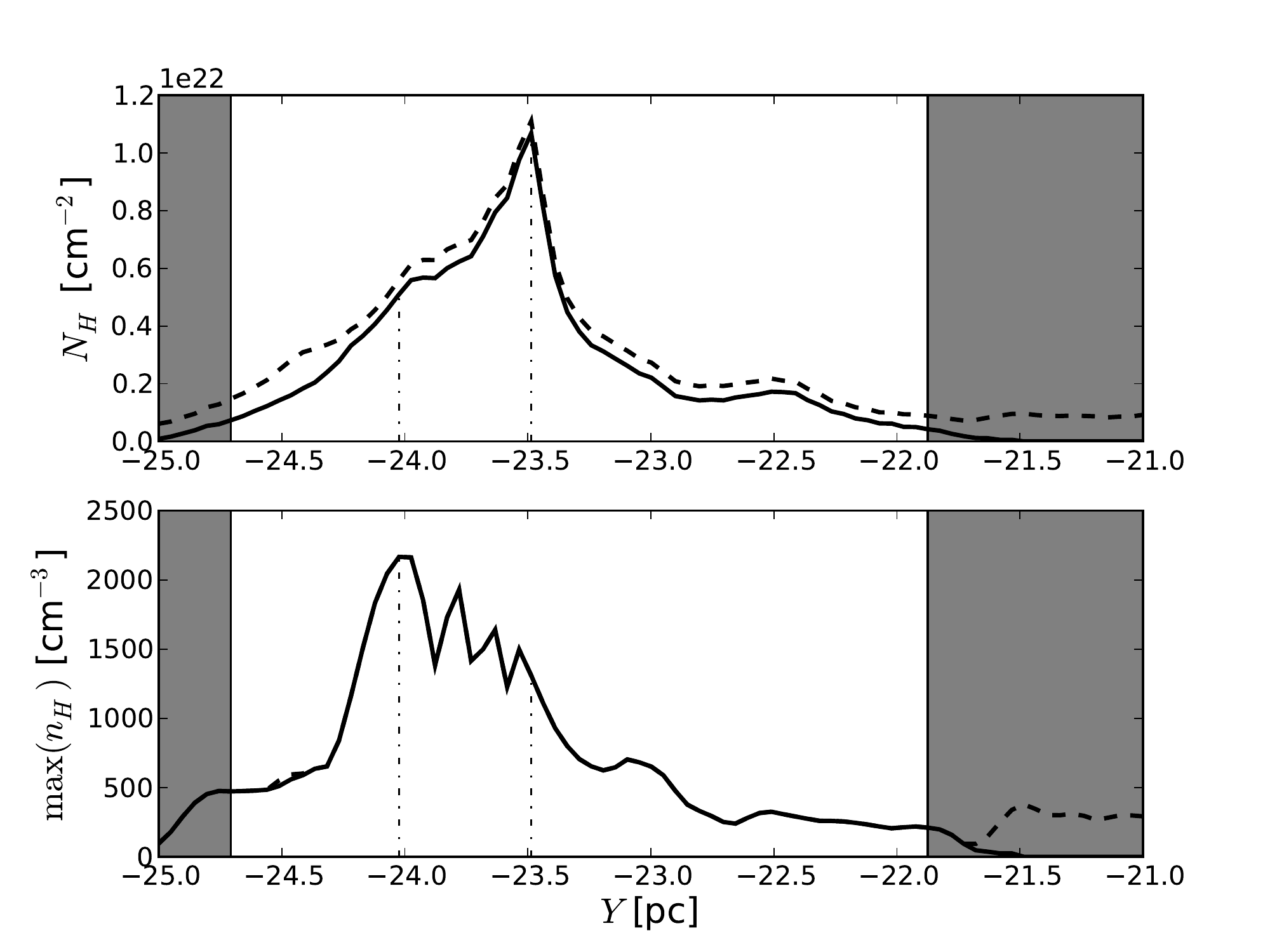}
      \caption{Total gas column densities along the $X$ axis within the {\bf AB} strip under study (top plot) and maximum gas density $\max{(n_H)}$ on the same lines of sight (bottom plot). Shown are the column densities over the full 50 pc lines of sight along the $X$ axis (dashed line) and the column densities for the overdense regions $n_H\geqslant n_0$ shown on Fig.~\ref{Fig_D_T_Clump} (solid line). Grey areas mark lines of sight for which less than 50\% of the mass is in the overdense region. The dash-dotted lines mark the positions of the maxima of $N_H$ and $\max{(n_H)}$, which are separated by 0.55~pc.}
         \label{Fig_coldens_comparison}
   \end{figure}

It is this subset of the simulated cube, shown on Fig.~\ref{Fig_D_T_Clump}, which is the focus of our study, and from which we extract one-dimensional density profiles. As Fig.~\ref{Fig_coldens_comparison} shows, this subset indeed contains most of the gas on the lines of sight within the {\bf AB} strip~: Except in the outermost regions where $N_H\leqslant 1.5~10^{21}~\textrm{cm}^{-2}$, column densities along $X$ over that region represent more than half the total column densities over the full 50 pc lines of sight. Fig.~\ref{Fig_coldens_comparison} also shows that, whithin the {\bf AB} strip, the peaks of $N_H$ and $\max{(n_H)}$ are still separated, by about 0.55~pc.

   \begin{figure}
   \centering
   \includegraphics[width=10cm]{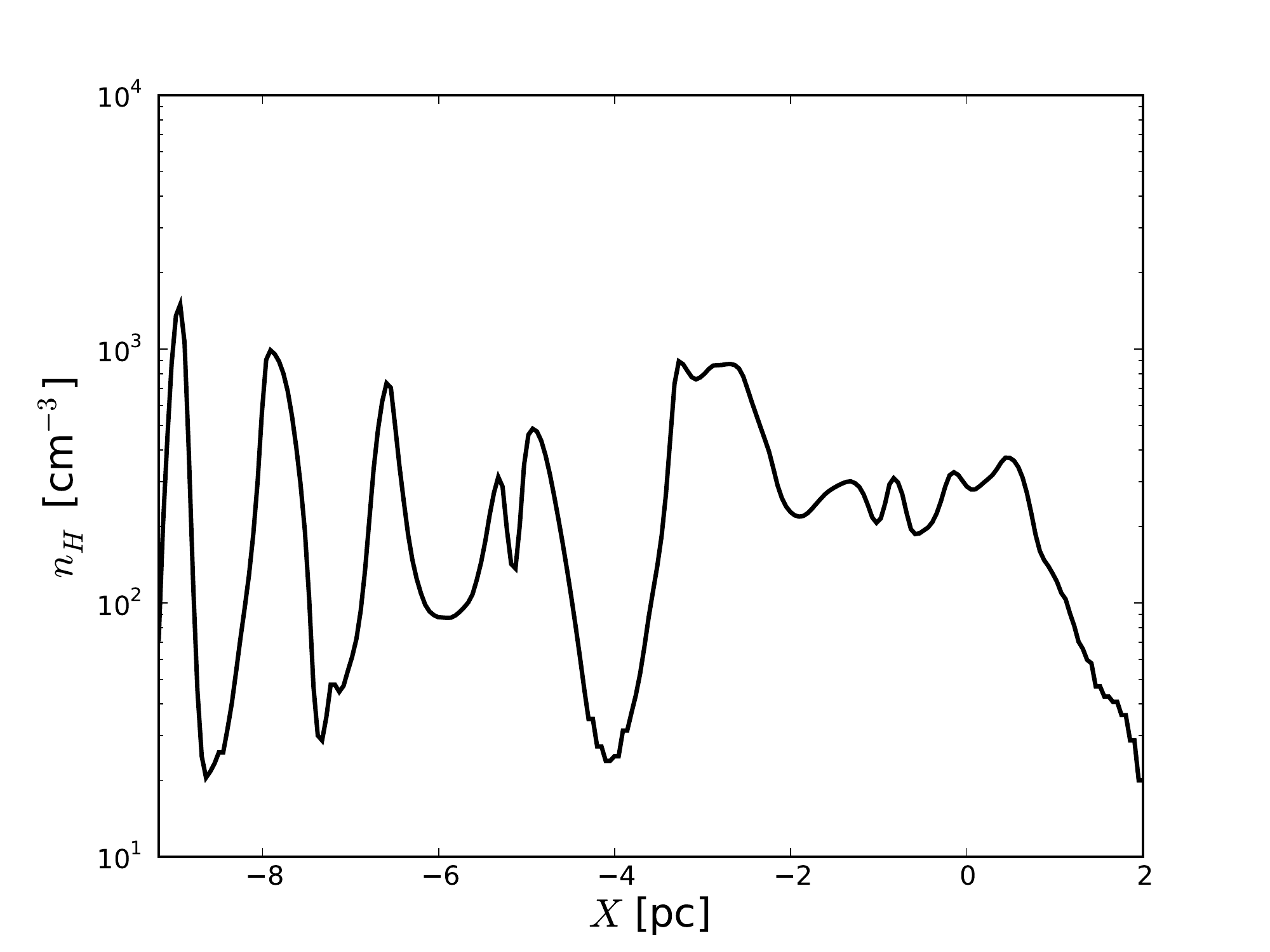}
      \caption{Example of a density profile used in the PDR code. This is the profile extracted at the location of the dashed line on Fig.~\ref{Fig_D_T_Clump}.}
         \label{FigSightLine}
   \end{figure}

As explained in section~\ref{sec:method}, we consider two ($p=2$) possible directions for the one-dimensional density profiles extracted from the 2D subset, namely those parallel to the $X$ or $Y$ axis. The dashed line on Fig.~\ref{Fig_D_T_Clump} marks the location of such an extracted profile, which is shown on Fig.~\ref{FigSightLine}. Note that we consider profiles to be connectedly overdense, which means that, along each line of sight, we may extract several profiles separated by underdense regions $n_H< n_0$. Such is the line of sight parallel to the $X$ axis located at $Y=-23~\mathrm{pc}$, for instance.

\begin{table}
\caption{Properties of the 156 one-dimensional profiles extracted parallel to the $X$ axis. Listed are the minimum, maximum and ensemble average values for the size, average density, column density, visual extinction (corresponding to $N_H$ via Eq.~\ref{eq:1}), density-weighted average temperature and line-of-sight velocity dispersion. For that last quantity, the minimum value is not shown, as it is too small to be meaningful.}
\label{table:0}
\centering
\begin{tabular}{c c c c}
\hline\hline          
Parameter $(F)$ & $\min{(F)}$ & $\max{(F)}$ & $\left<{F}\right>$ \\
\hline
Size $L$ [pc] &  0.15 & 11.2 & 2.18 \\
Density $\left<n_H\right>$ [cm$^{-3}$] &  20 & 571 & 155 \\
Column density $N_H$ [$10^{20}$ cm$^{-2}$] & $0.117$ & $107$ & $11.9$ \\
Visual extinction $A_V$ & $6.3~10^{-3}$ & 5.7 & 0.64 \\
Temperature $\overline{T}$ [K] & 22 & 924 & 88 \\
Velocity dispersion $\sigma_X$ [km.s$^{-1}$] & $-$ & 1.9 & 0.50 \\
\hline    
\end{tabular}
\end{table}

\begin{table}
\caption{Same as Table~\ref{table:0} but for the 291 one-dimensional profiles extracted parallel to the $Y$ axis.}
\label{table:0bis}
\centering
\begin{tabular}{c c c c}
\hline\hline          
Parameter $(F)$ & $\min{(F)}$ & $\max{(F)}$ & $\left<{F}\right>$ \\
\hline
Size $L$ [pc] &  0.24 & 2.78 & 1.17 \\
Density $\left<n_H\right>$ [cm$^{-3}$] &  22 & 655 & 188 \\
Column density $N_H$ [$10^{20}$ cm$^{-2}$] & $0.165$ & $31.6$ & $6.39$ \\
Visual extinction $A_V$ & $8.8~10^{-3}$ & 1.7 & 0.34 \\
Temperature $\overline{T}$ [K] & 21 & 464 & 56 \\
Velocity dispersion $\sigma_Y$ [km.s$^{-1}$] & $-$ & 1.51& 0.37 \\
\hline    
\end{tabular}
\end{table}

We thus extract 447 density profiles, of which 156 are parallel to the $X$ axis and 291 are parallel to the $Y$ axis. Their statistical properties are summarized in Tables~\ref{table:0} and~\ref{table:0bis}, emphasizing the large dynamic range they sample in column densities ($\sim10^4$) and mean densities ($\sim 30$). The large values in density-weighted temperatures correspond to density profiles that never much deviate from $n_0=20~\mathrm{cm}^{-3}$. For each of these 447 profiles, we run the PDR code assuming identical illumination on both sides. Since one of the objectives of this paper is to assess the effect of realistic density distributions along the line of sight on the chemical composition of interstellar clouds, we also apply the PDR code on a homogeneous reference model for each extracted profile. We specify this model, called {\tt uniform} in the following, as having the same mean density $\left<n_H\right>$ and total visual extinction $A_V$ as the inhomogeneous model, which we dub {\tt los} from now on. Unless otherwise specified, results presented in the next section refer to these {\tt los} models. The setup for all runs is detailed in appendix~\ref{sec:PDR-wdp} and their post-processing is described in appendix~\ref{sec:postprocessing}.

\section{Results}
\label{sec:full}

\subsection{Temperature comparison}

\begin{figure}
   \centering
   \includegraphics[width=9.9cm]{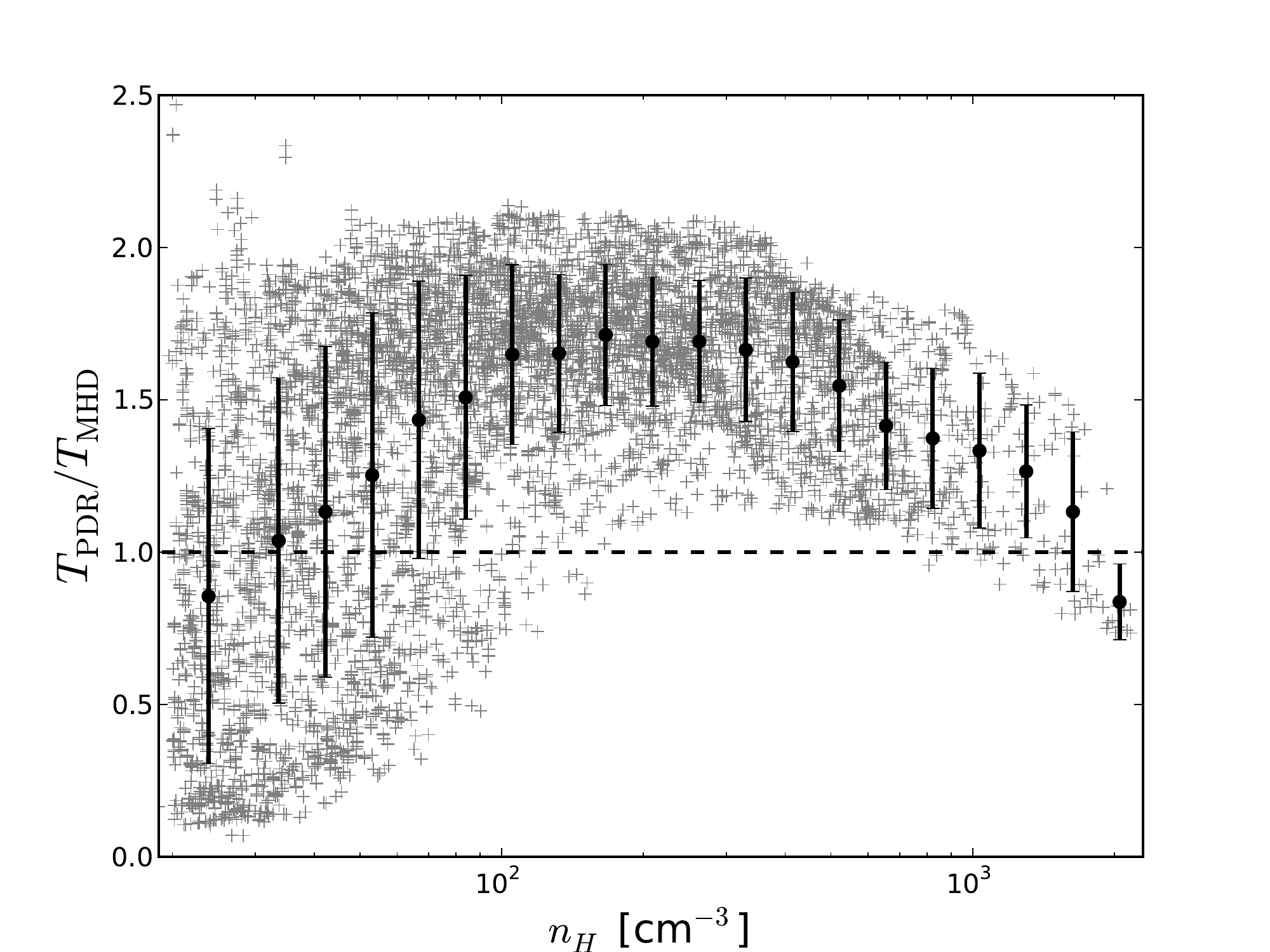}
      \caption{Ratio of the gas temperatures $T_\mathrm{PDR}/T_\mathrm{MHD}$ versus total gas density $n_H$. Grey crosses show all points, while black circles are average values over density bins in logarithmic scale, with error bars standing for $\pm 1\sigma$. The dashed line corresponds to $T_\mathrm{PDR}/T_\mathrm{MHD}=1$.}
      \label{fig:temperature_ratio_dens}
   \end{figure}

The PDR code and the MHD simulation both treat thermal balance, so that we have two estimates of the gas temperature, which we can compare : Fig.~\ref{fig:temperature_ratio_dens} shows the ratio $r=T_\mathrm{PDR}/T_\mathrm{MHD}$ of the gas temperature $T_\mathrm{PDR}$ output by the PDR code to the gas temperature $T_\mathrm{MHD}$ computed in the MHD simulation, at every point in the subset under study, versus the total gas density $n_H$ at that point. Average ratios $\left<r\right>$ in selected density bins are also shown. It appears quite clearly that $\left<r\right>$ is close to 1, with $0.3\lesssim \left<r\right> \lesssim 2.0$ over the whole range of densities. 

The fact that $T_\mathrm{PDR}\sim T_\mathrm{MHD}$ actually comes as a pleasant surprise, considering the differences between the PDR and MHD computations : while the former is 1D, steady-state, and includes cooling via the infrared and submillimeter lines from atomic and molecular species, especially H$_2$ transitions \citep{lepetit06}, the latter is 3D, dynamical, and only includes cooling via the fine structure lines of CII [158$\mu$m] and OI [63$\mu$m], as well as the recombination of electrons with ionized PAHs. This suggests that unless a very precise knowledge of the temperature is needed, it is probably not necessary, at least as a first approximation and in the range of densities and temperatures probed here, to refine the details of cooling processes in our MHD simulations, as the simple cooling function currently used already yields gas temperatures close to those found using the more detailed processes of the PDR code. Similar conclusions were reached by \citet{glover10}.

\subsection{Chemical structure and comparison to observations}

\begin{figure}
   \centering
   \includegraphics[width=8.7cm]{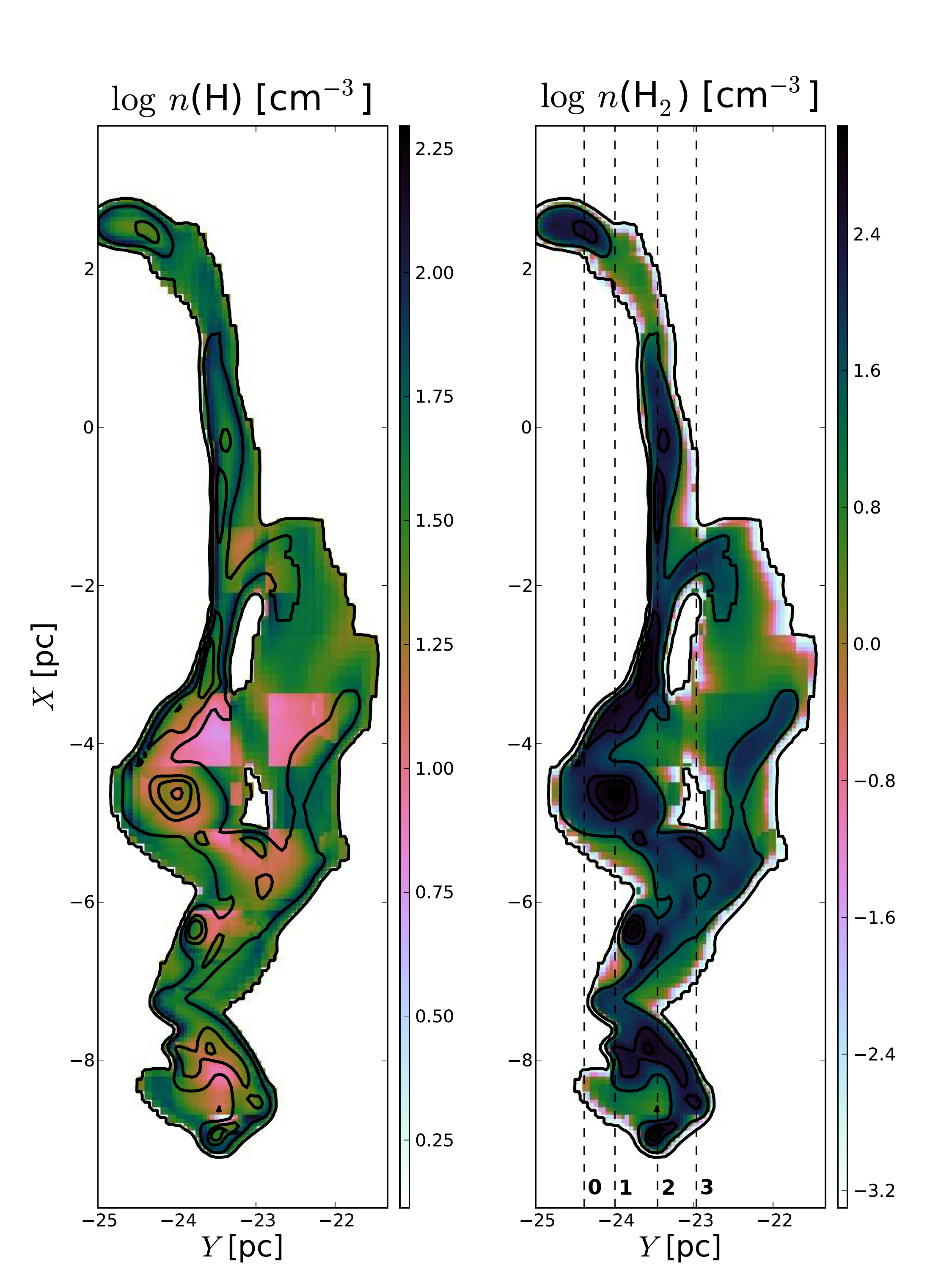}
   \includegraphics[width=8.7cm]{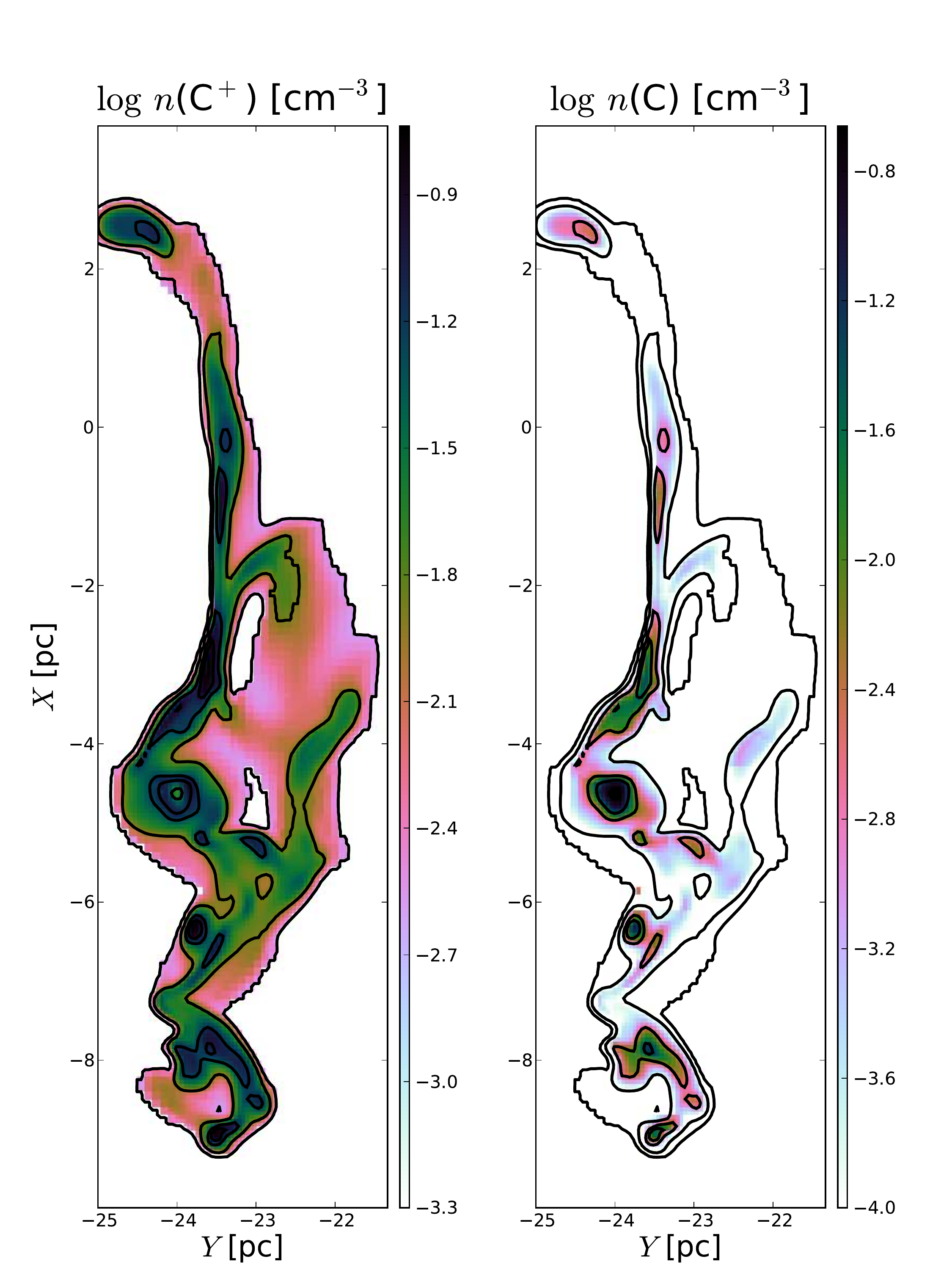}
      \caption{H (top left), H$_2$ (top right), C$^+$ (bottom left) and C (bottom right) abundances for the {\tt los} models. Contours mark total gas densities 20, 100, 500, 1000 and 2000~$\mathrm{cm}^{-3}$. C$^+$ and C abundance maps are clipped below $2~10^{-4}~\mathrm{cm}^{-3}$ and $10^{-4}~\mathrm{cm}^{-3}$, respectively. Lines of sight {\bf 0}, {\bf 1}, {\bf 2} and {\bf 3} on the $n(\mathrm{H}_2)$ map refer to the discussion in the text.}
      \label{fig:species_abundances}
   \end{figure}

\begin{figure}
   \centering
   \includegraphics[width=8.7cm]{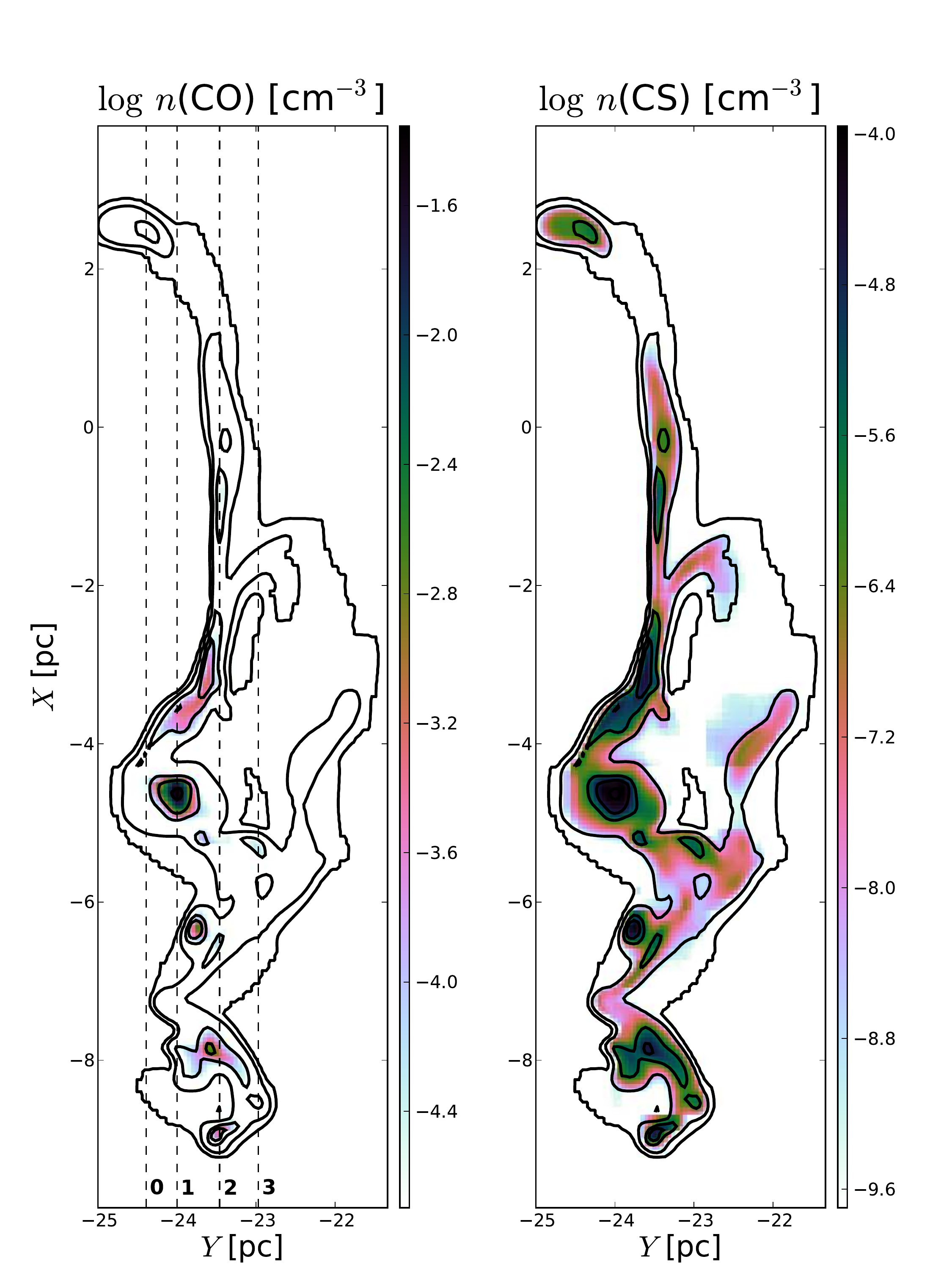}
   \includegraphics[width=8.7cm]{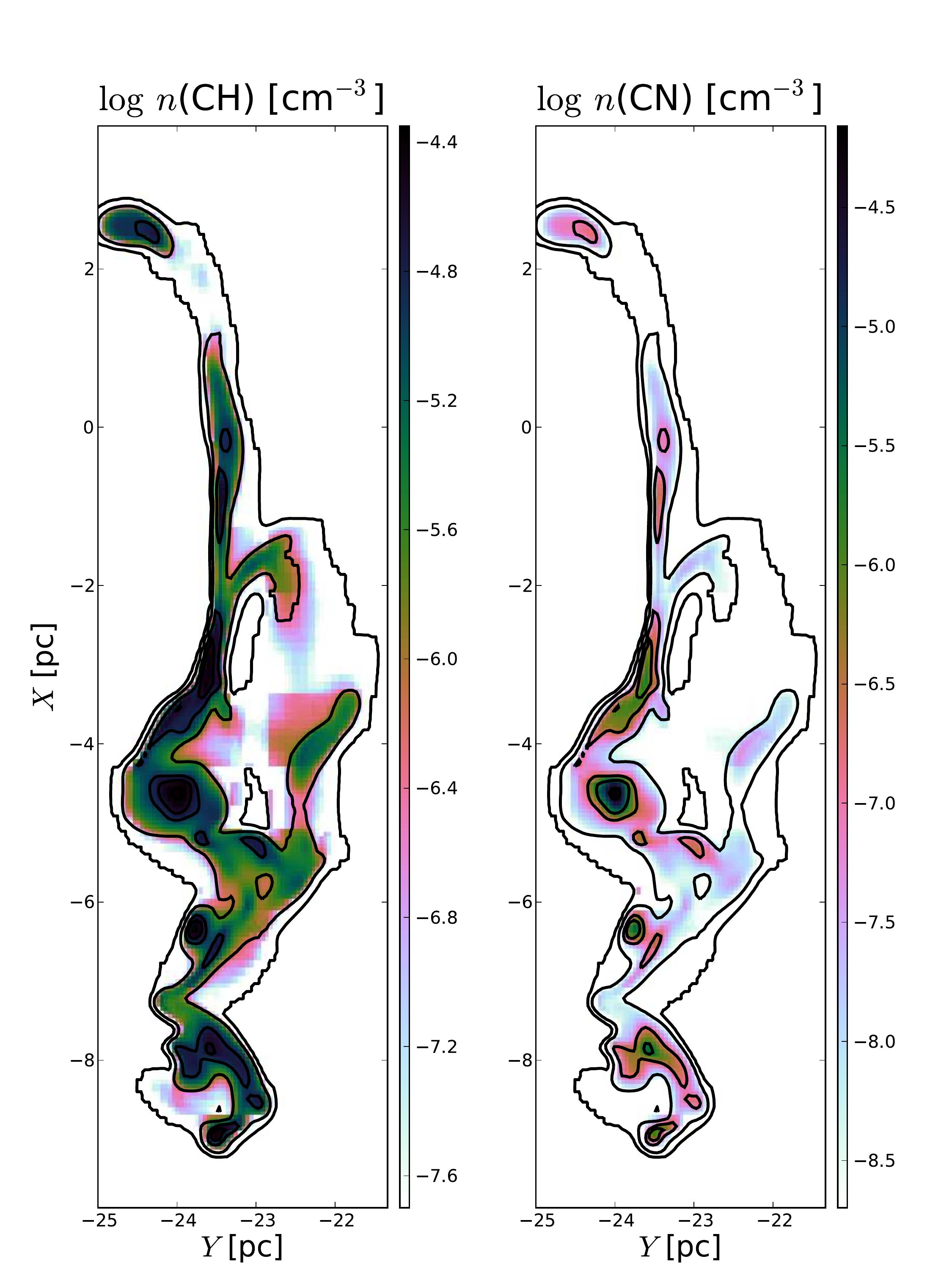}
      \caption{Same as Fig.~\ref{fig:species_abundances} but for CO (top left), CS (top right), CH (bottom left) and CN (bottom right). Abundance maps are clipped below $2~10^{-5}~\mathrm{cm}^{-3}$ for CO, $2~10^{-10}~\mathrm{cm}^{-3}$ for CS, $2~10^{-8}~\mathrm{cm}^{-3}$ for CH and $2~10^{-9}~\mathrm{cm}^{-3}$ for CN.}
      \label{fig:species_abundances_2}
   \end{figure}

\begin{figure}
   \centering
   \includegraphics[width=9.5cm]{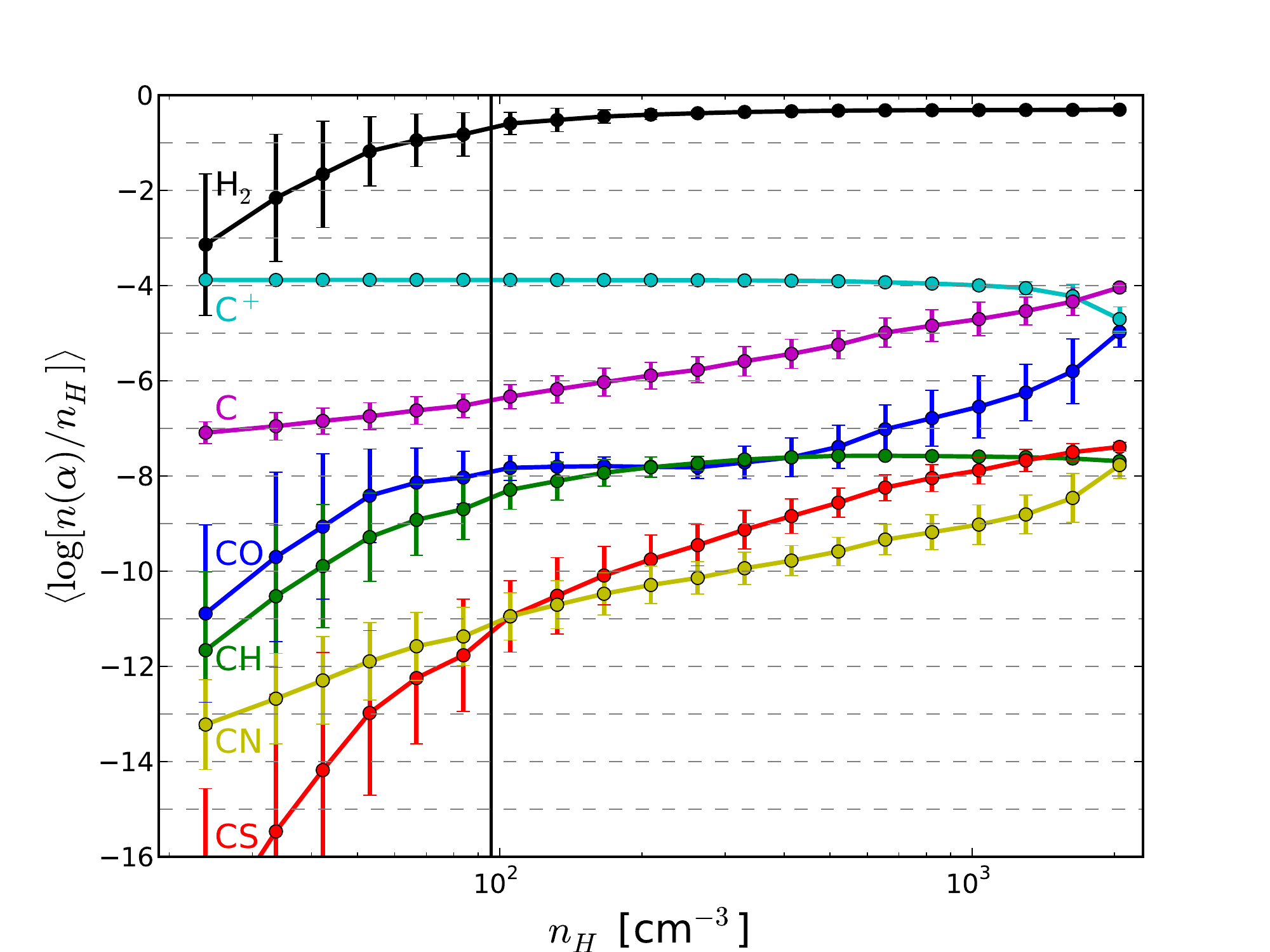}
      \caption{Abundances of H$_2$, C$^+$, C, CO, CH, CS and CN versus total gas density $n_H$. Data points are averaged in the same $n_H$ bins as on Fig.~\ref{fig:temperature_ratio_dens}. The vertical line marks the position of the average molecular transition where $\left<f_{\mathrm{H}_2}\right>=2\left<n({\mathrm{H}_2})/n_H\right>=1/2$.}
      \label{fig:dtracers}
   \end{figure}

\begin{table}
\caption{Typical threshold abundances $X_\alpha$ used in Figs.~\ref{fig:species_abundances} and \ref{fig:species_abundances_2}.}
\label{table:Xs}
\centering
\begin{tabular}{l l l}
\hline\hline          
C$^+$ : $10^{-5}$ & CO : $10^{-6}$ & CH : $10^{-9}$\\
C : $5~10^{-6}$ & CS : $10^{-11}$ & CN : $10^{-10}$\\
\hline    

\end{tabular}
\end{table}
The spatial distributions of H, H$_2$, C$^+$, C, CO, CS, CH and CN in the simulation subset are shown on Figs.~\ref{fig:species_abundances} and \ref{fig:species_abundances_2}. In these figures, points were clipped where the density $n(\alpha)$ of species $\alpha$ was below $n_{0,\alpha}=X_\alpha n_0$, with $X_\alpha$ a typical threshold abundance for the detection of that species (see Table~\ref{table:Xs}). Although shadowing effects remain (for instance on the atomic hydrogen map), these figures show how some species (e.g. CO, CN) trace denser gas than others (C$^+$, CH). This appears more clearly when plotting these abundances, averaged over density bins, versus total gas density $n_H$ (Fig.~\ref{fig:dtracers}) : C$^+$ traces gas uniformly up to $n_H\gtrsim 10^3~\mathrm{cm}^{-3}$, while CO starts rising up at $n_H\gtrsim 250~\mathrm{cm}^{-3}$, right about where CH flattens out. This break in the slope for CO occurs after the molecular transition $\left<f_{\mathrm{H}_2}\right>=2\left<n({\mathrm{H}_2})/n_H\right>=1/2$, which is at $n_H\simeq 100~\mathrm{cm}^{-3}$. Considering the abundances of C and C$^+$, this means that a significant fraction of the molecular gas (i.e. where hydrogen is mostly in the form of H$_2$) is better traced by C and C$^+$ than by CO. This "dark molecular gas" fraction is the subject of subsection~\ref{sec:darkgas}. CH, on the other hand, nicely follows H$_2$~\citep{sheffer08}. To complete the picture, C and CN have a similar slope throughout the density range, although CN seems to break away slightly at $n_H\gtrsim 10^3~\mathrm{cm}^{-3}$, to follow CO.
   
\begin{figure}
   \begin{center}
   \includegraphics[width=7.8cm]{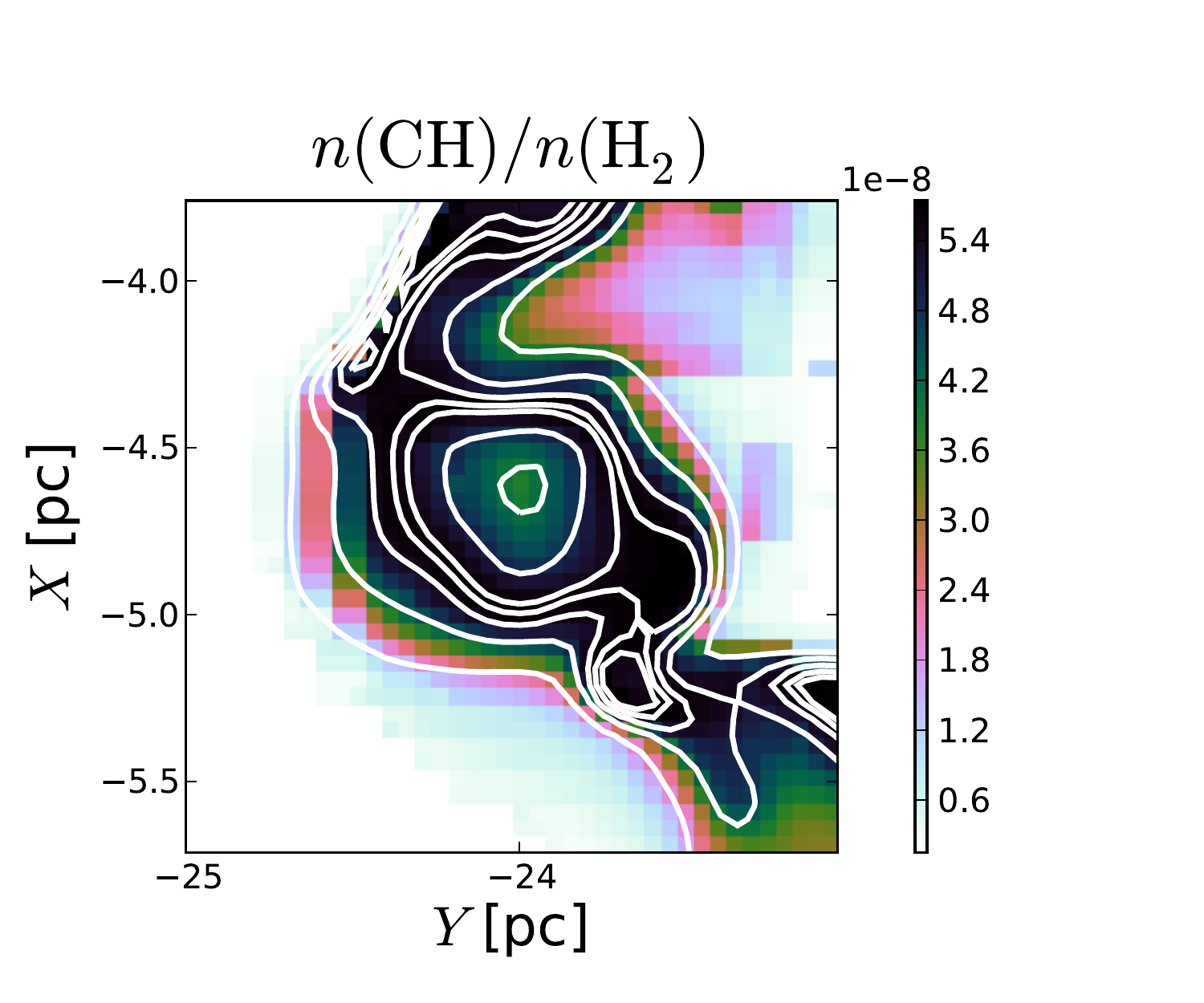}
   \includegraphics[width=7.8cm]{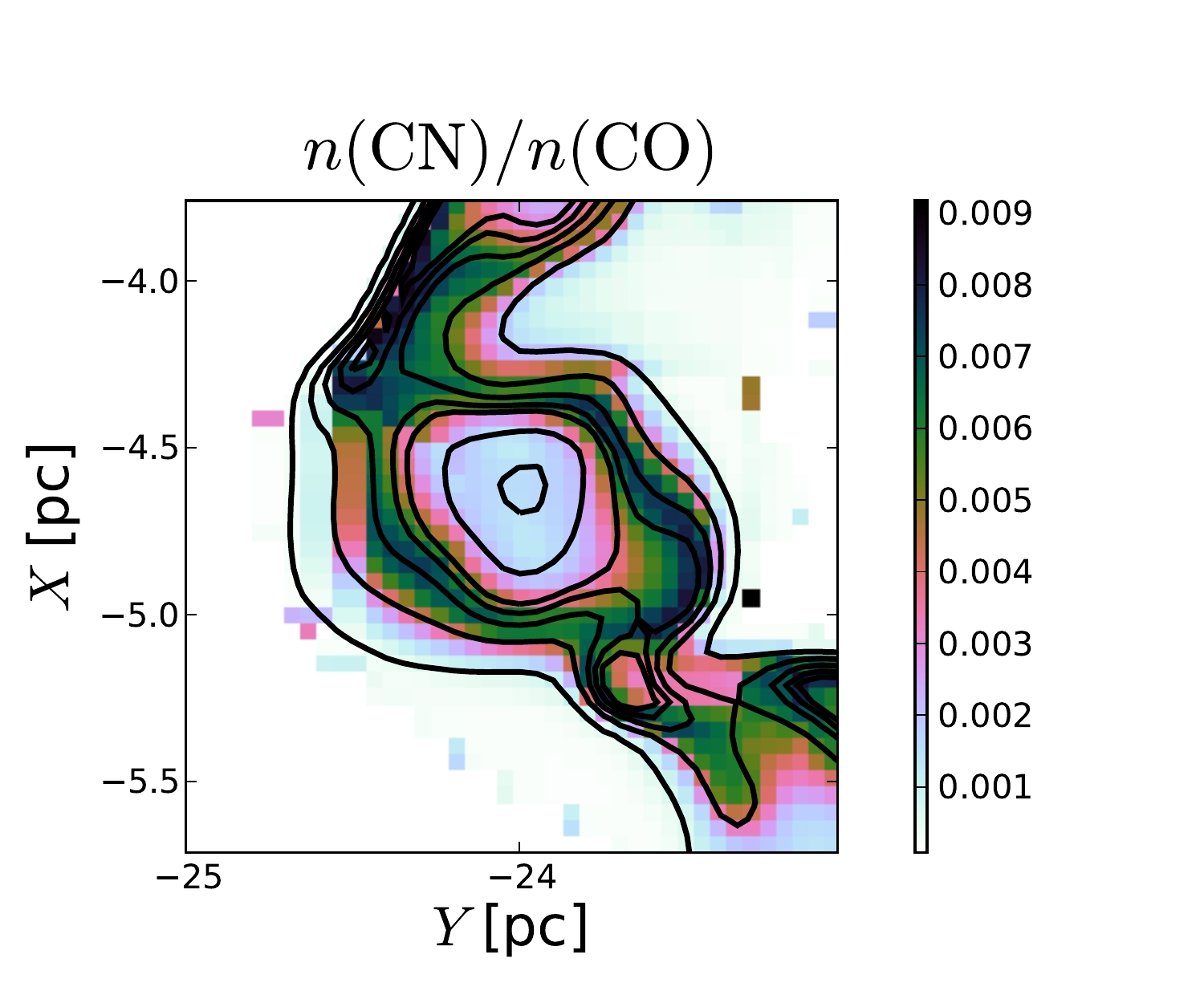}
   \includegraphics[width=7.8cm]{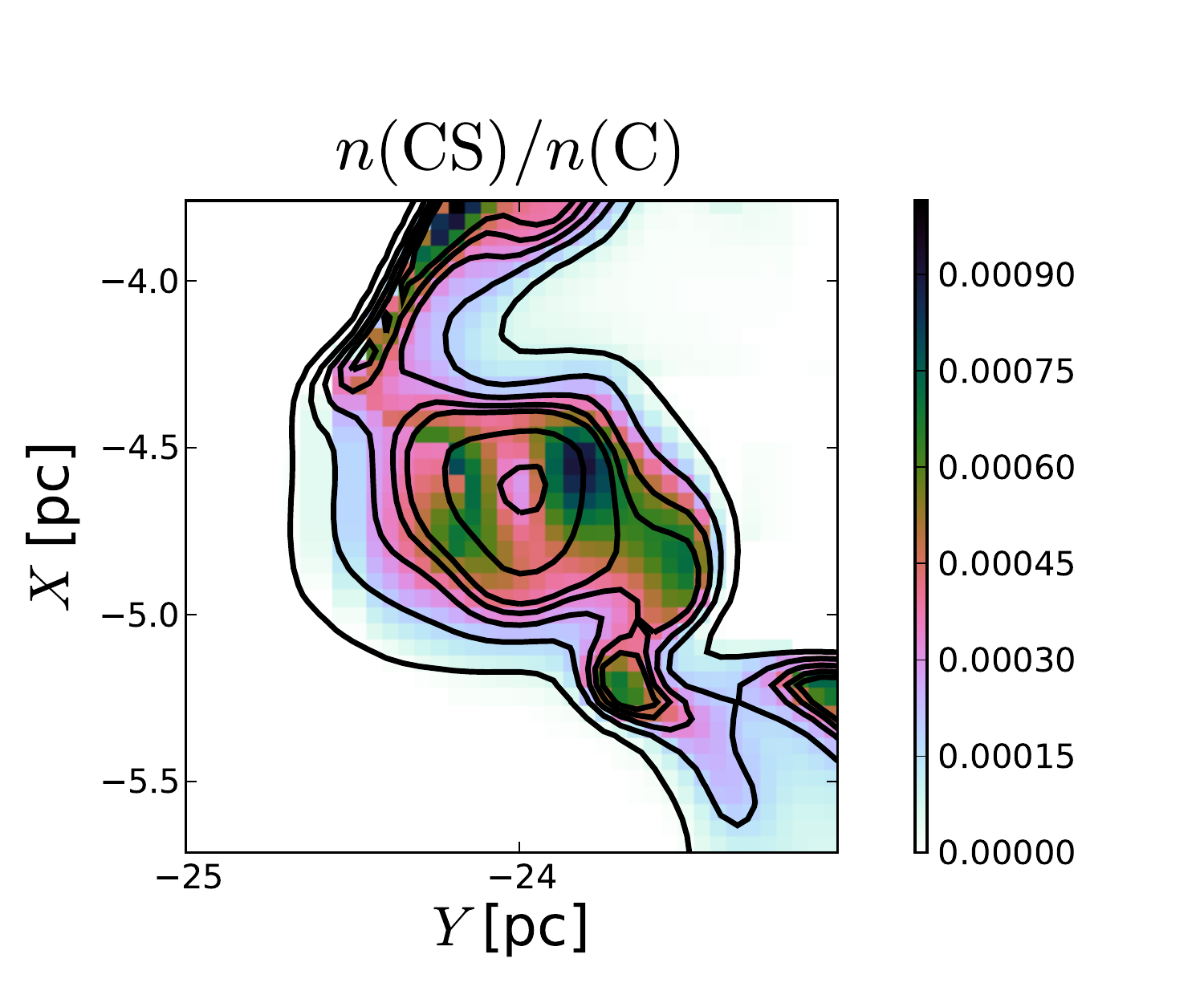}
   \end{center}
      \caption{Abundance ratios $n(\mathrm{CH})/n(\mathrm{H_2})$ (top), $n(\mathrm{CN})/n(\mathrm{CO})$ (middle) and $n(\mathrm{CS})/n(\mathrm{C})$ (bottom) in the vicinity of the total gas density peak, located at $(X\simeq-4.7~\mathrm{pc},Y\simeq -24~\mathrm{pc})$. Contours mark total gas densities $n_H$ of 100, 200, 300, 400, 500, 1000 and 2000~$\mathrm{cm}^{-3}$.}
      \label{fig:core_abundance_ratios}
   \end{figure}

To be more precise on these apparent correlations (CH vs. H$_2$, CS vs. C and CN vs. CO), we show, on Fig.~\ref{fig:core_abundance_ratios}, abundance ratios for these similarly distributed species in the region where they are all significantly present, that is the cloudlet at $(X\simeq-4.7~\mathrm{pc},Y\simeq -24~\mathrm{pc})$ (see Figs.~\ref{fig:species_abundances} and \ref{fig:species_abundances_2}). We can see that $n(\mathrm{CH})/n(\mathrm{H_2})$ and $n(\mathrm{CN})/n(\mathrm{CO})$ have a similar "ringlike" behaviour, rising to maximum values $n(\mathrm{CH})/n(\mathrm{H_2})\sim 5.5~10^{-8}$ and $n(\mathrm{CN})/n(\mathrm{CO})\sim 10^{-2}$ at total gas densities $n_H\sim 400-500~\mathrm{cm}^{-3}$, then falling at higher densities, to $n(\mathrm{CH})/n(\mathrm{H_2})\sim 4~10^{-8}$ and $n(\mathrm{CN})/n(\mathrm{CO})\sim 2~10^{-3}$, respectively. For $n(\mathrm{CS})/n(\mathrm{C})$, there is a slight loss of azimuthal symmetry, but the overall trend is the same, although maximum values of  $\sim 10^{-3}$ are reached at larger densities $n_H\sim 1000~\mathrm{cm}^{-3}$ before falling to $\sim 3~10^{-4}$ at the peak. 

\begin{figure}
   \centering
   \includegraphics[width=9.2cm]{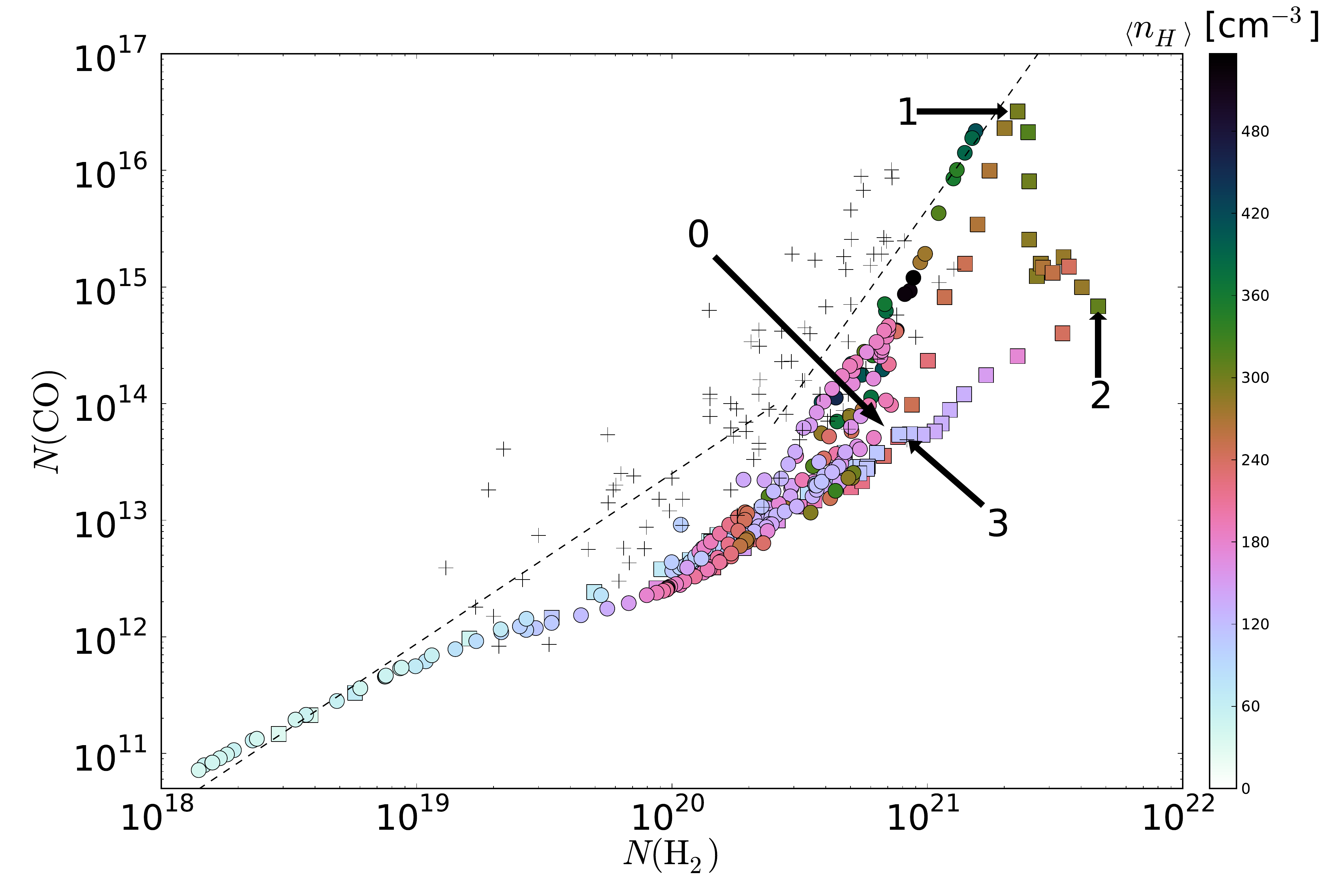}
      \includegraphics[width=9.2cm]{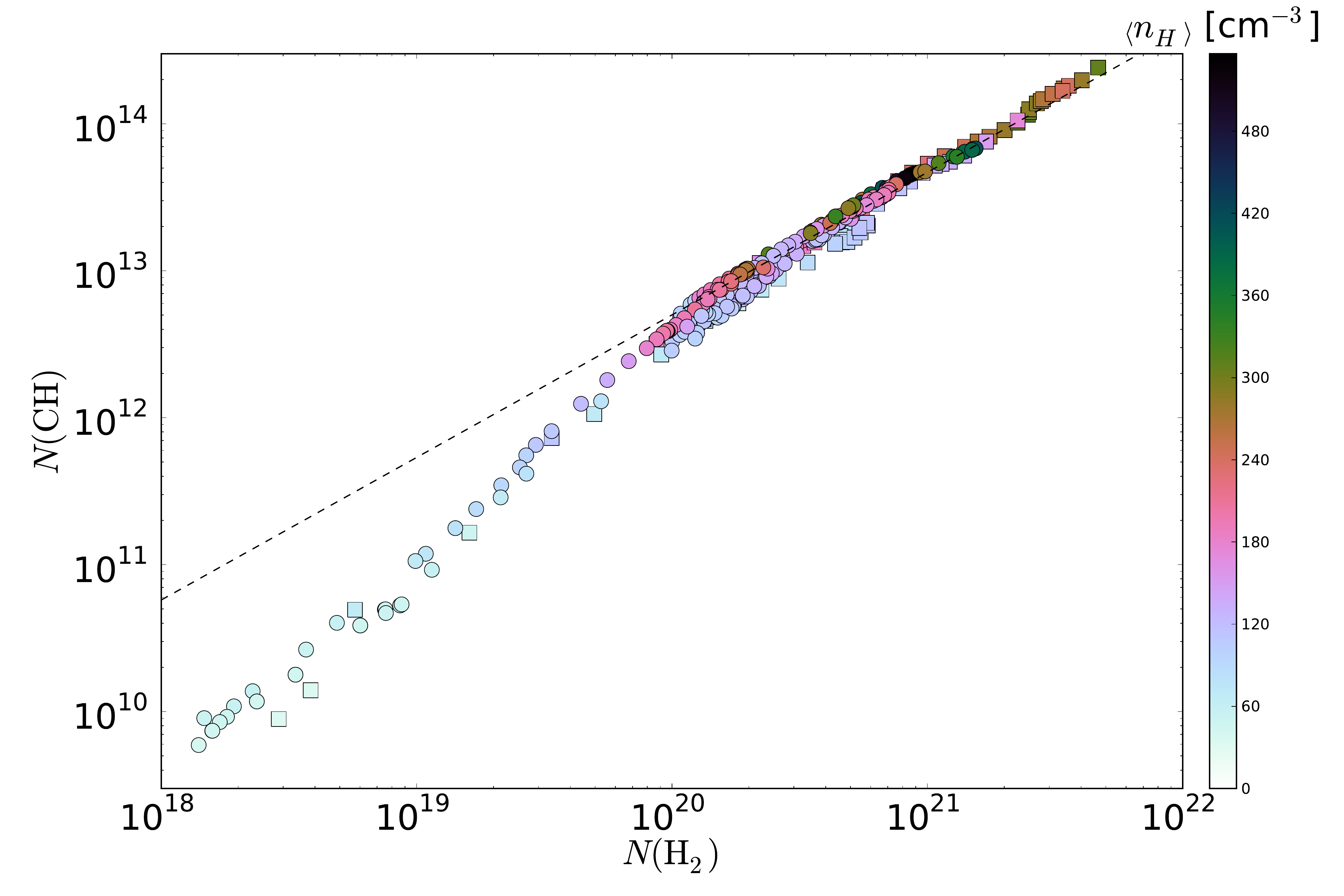}
         \includegraphics[width=9.2cm]{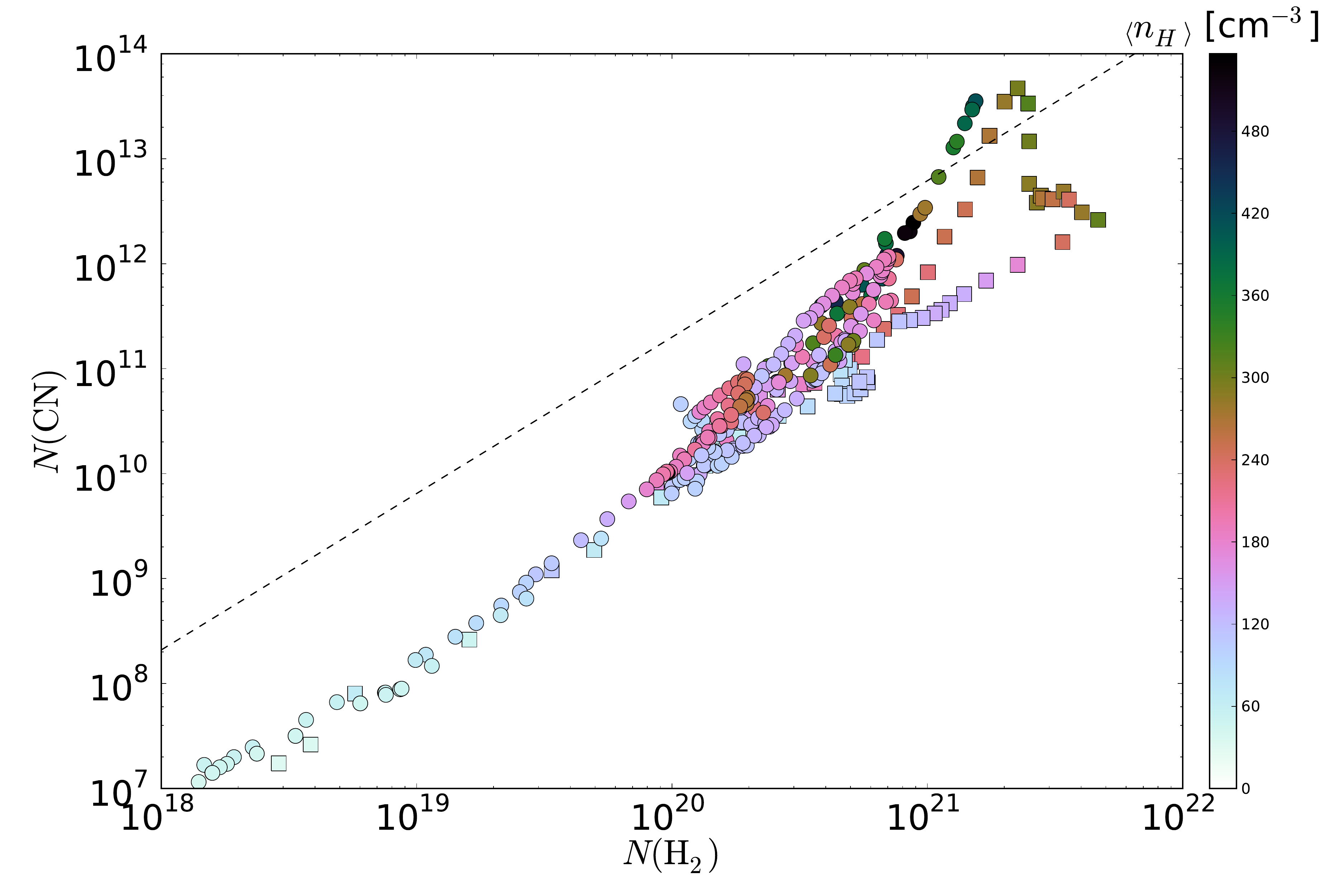}
      \caption{Column densities of CO (top), CH (middle) and CN (bottom) versus column densities of $\mathrm{H_2}$, in the {\tt los} models. Circles correspond to lines of sight parallel to $Y$ and squares to lines of sight parallel to $X$. Their colours reflect the mean gas density $\left<n_H\right>$ on these lines of sight. Plus signs on the top (CO) plot stand for observational data points~\citep{sheffer08,crenny04,pan05,lacour05,rachford02,rachford09,snow08}. The dashed lines are power-law fits from \citet{sheffer08}. The lines of sight parallel to $X$ marked {\bf 0}, {\bf 1}, {\bf 2} and {\bf 3} on the top panel are the same as on Figs.~\ref{fig:species_abundances} and \ref{fig:species_abundances_2}.}
      \label{fig:CO-H2}
   \end{figure}

The comparison to observational data requires us to work not with densities but with column densities, which are what observers have access to. To his end, we compute column densities for CO, CH and CN on each line of sight parallel to the $X$ or $Y$ axis and plot them versus those of H$_2$ (Fig.~\ref{fig:CO-H2}), to match the observational data plots in \citet{sheffer08}. We use a colour scheme to specify the mean gas density $\left<n_H\right>$ on the line of sight, and plot separately data points corresponding to the long lines of sight parallel to $X$ (squares) and to the shorter lines of sight parallel to $Y$ (circles). Fits to observational data derived by \citet{sheffer08} are shown as dashed lines, and on the CO panel, we also plot actual data points from that same paper.

Regarding CO, our data shows a significant deficit around $N(\mathrm{H}_2)\sim 10^{20}~\mathrm{cm}^{-2}$, which may be a general issue with PDR chemistry computations~\citep{sonnentrucker07}. However, the agreement with the observational fit gets better, both in values and in slope, for $N(\mathrm{H}_2)\gtrsim 2~10^{20}~\mathrm{cm}^{-2}$. The slope found is $\mathrm{d}[\log N(\mathrm{CO})]/\mathrm{d}[\log N(\mathrm{H}_2)]\simeq 5.2$, while the observational fit yields $3.07\pm 0.73$. On the longer lines of sight, we can see a "loop" structure which needs to be understood. It is easily identified with lines of sight between $Y=-24.5~\mathrm{pc}$ and $Y=-23~\mathrm{pc}$ (positions {\bf 0} to {\bf 3} on the H$_2$ and CO maps from Figs.~\ref{fig:species_abundances} and \ref{fig:species_abundances_2}). To be more accurate, following the loop clockwise corresponds to scanning lines of sight parallel to $X$ from {\bf 0} to {\bf 3}, with turnovers at positions {\bf 1} and {\bf 2}. Considering the H$_2$ and CO maps alongside Fig.~\ref{fig:CO-H2} helps to understand this "loop". Firstly, lines of sight from {\bf 0} to {\bf 1} basically intercept just one dense structure with both H$_2$ and CO, so that we have a very similar behaviour to that of the short lines of sight parallel to $Y$. Secondly, lines of sight from {\bf 1} to {\bf 2} pass through many less dense CO structures, with a lot of molecular hydrogen in between, leading to the sharp drop in the $N(\mathrm{CO})$ vs. $N(\mathrm{H}_2)$ relation. At a given $N(\mathrm{CO})$, the difference $\Delta N(\mathrm{H}_2)$ between H$_2$ column densities in the branches {\bf 0}-{\bf 1} and {\bf 1}-{\bf 2} thus represents the "dark gas" (see subsection~\ref{sec:darkgas}). Finally, lines of sight from {\bf 2} to {\bf 3} barely have any CO and contain less and less H$_2$ as we approach position {\bf 3}, where we return to a situation similar to that at position {\bf 0}. The split between the two branches is definitely related to the mean gas density on the line of sight, the upper branch having $\left<n_H\right>\sim 250~\mathrm{cm}^{-3}$, the lower one having $\left<n_H\right>\sim 150~\mathrm{cm}^{-3}$. The overall deficit in CO suggests that we may not shield it enough from the ambient UV field, a hypothesis which we discuss in section~\ref{sec:conclusions}.

The behaviour of $N(\mathrm{CH})$ with respect to $N(\mathrm{H}_2)$ (Fig.~\ref{fig:CO-H2} - middle panel) is in remarkable agreement with the observational data fit by \citet{sheffer08}. Only for $N(\mathrm{H}_2)\lesssim 10^{20}~\mathrm{cm}^{-2}$ is there a discrepancy, and it is irrelevant, as there are no detections in that range, only upper limits.

Concerning CN (Fig.~\ref{fig:CO-H2} - bottom panel), we typically have a factor 10 deficit when comparing to observations. It appears that the reaction rate coefficient for the $\mathrm{CN}+\mathrm{N}\to \mathrm{C}+\mathrm{N}_2$ reaction in the KIDA database\footnote{\tt http://kida.obs.u-bordeaux1.fr/} may be too high (E. Roueff, private comm.), which could partly explain that deficit. 

\subsection{Comparison of the {\tt los} and {\tt uniform} models}

\begin{figure}
   \centering
   \includegraphics[width=9.2cm]{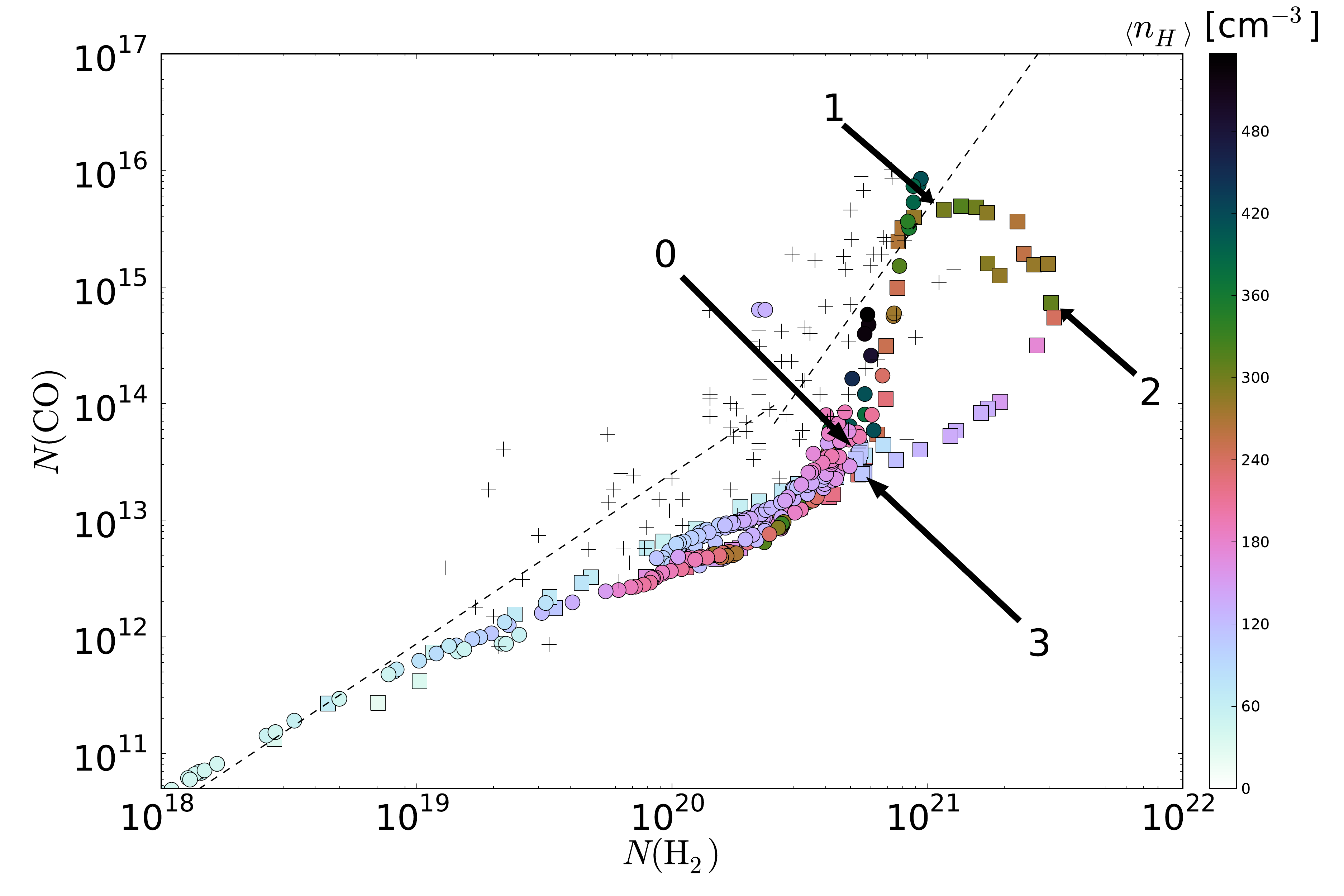}
   \includegraphics[width=9.2cm]{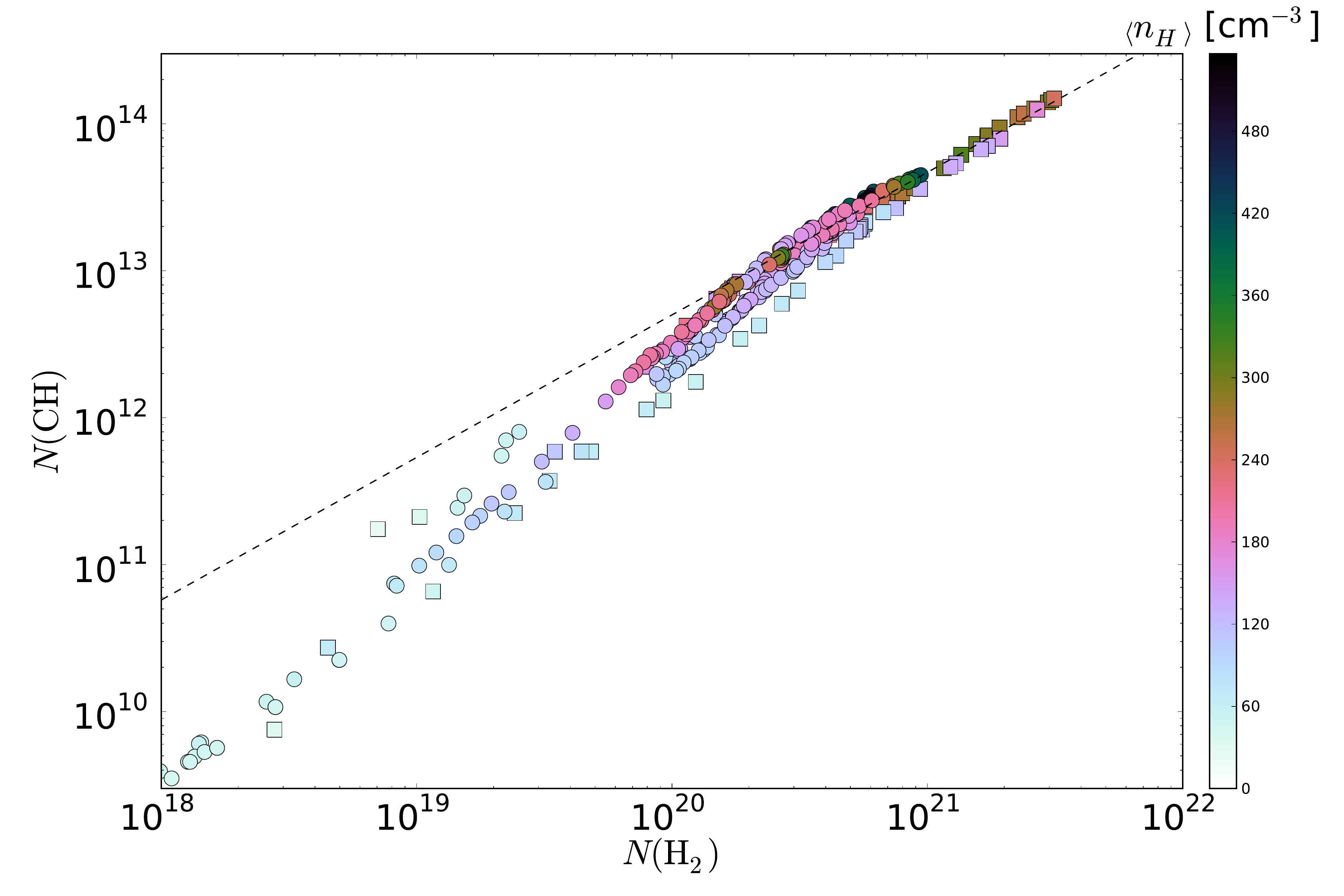}
   \includegraphics[width=9.2cm]{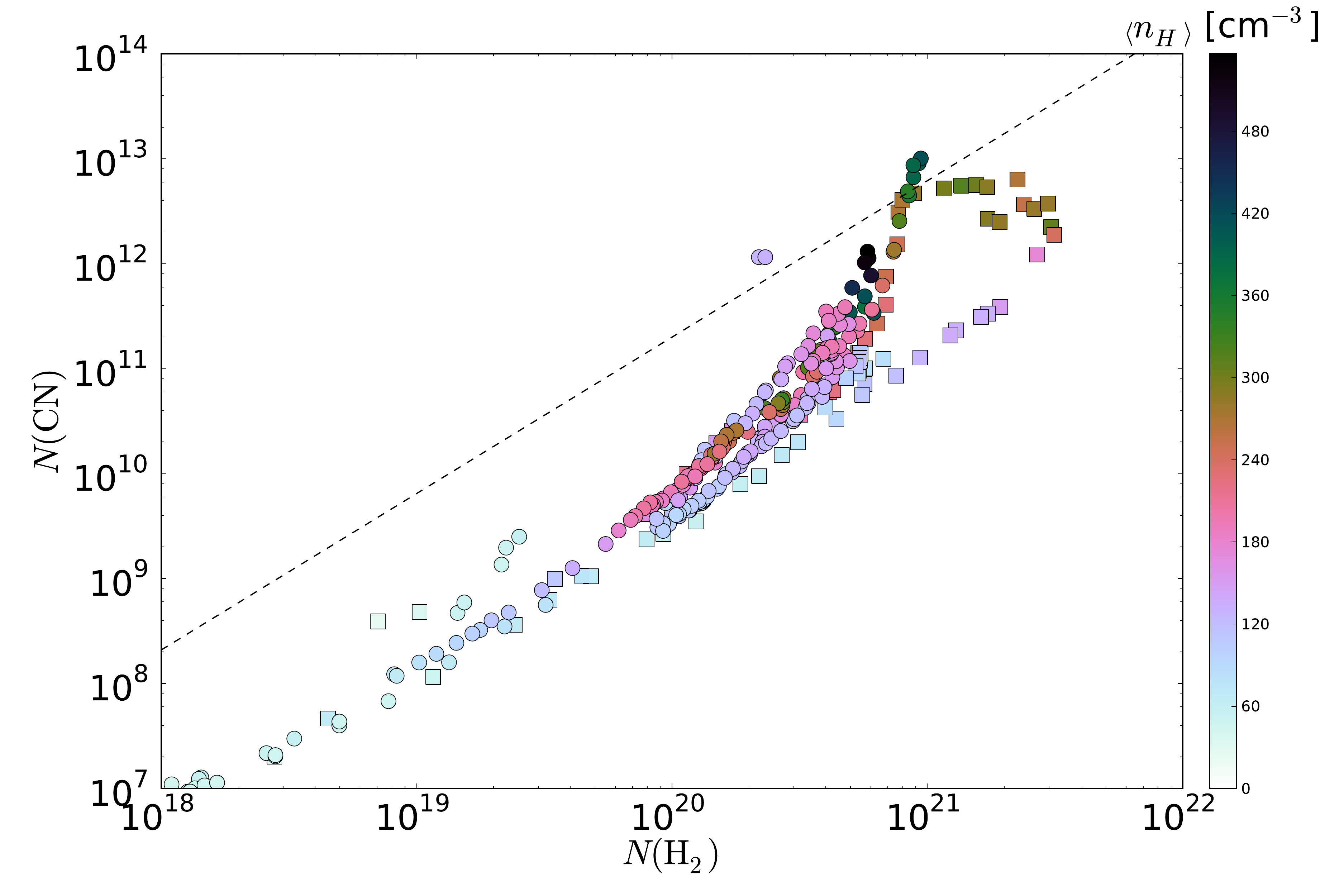}
         \caption{Same as Fig.~\ref{fig:CO-H2} but for the {\tt uniform} models.}
      \label{fig:CO-H2-uniform}
   \end{figure}

To assess the effects of taking into account density fluctuations, as opposed to the assumption of a homogeneous medium, usually made when modelling PDRs, we compute the column densities of CO and H$_2$ derived from the {\tt uniform} models, and plot them on Fig.~\ref{fig:CO-H2-uniform}. The behaviour at low $N(\mathrm{H}_2)$ is very similar to that in the {\tt los} models (Fig.~\ref{fig:CO-H2}), and the data suffer from the same deficit in CO around $N(\mathrm{H}_2)\sim 10^{20}~\mathrm{cm}^{-2}$. However, there is a definite difference at higher H$_2$ column densities, as the slope of the relation between both column densities is much steeper, $\mathrm{d}[\log N(\mathrm{CO})]/\mathrm{d}[\log N(\mathrm{H}_2)]\simeq 14$, than in the {\tt los} models and in observational data fits. This break occurs later, at $N(\mathrm{H}_2)\sim 5~10^{20}~\mathrm{cm}^{-2}$, and applies to the short lines of sight (parallel to $Y$ and parallel to $X$ between positions {\bf 0} and {\bf 1}) where there is essentially one structure in both H$_2$ and CO. However, the maximum column densities reached are slightly less than in the {\tt los} models by a factor $\sim 3$, for both CO and H$_2$. This shows the importance of taking into account density fluctuations along the line of sight when modelling PDRs.

For CH, the behaviour is very similar in the {\tt uniform} and {\tt los} models, although here also the maximum column densities reached are slightly less in the {\tt uniform} models, by a factor $\sim 2$. The observational fit is recovered at somewhat higher H$_2$ column densities ($2~10^{20}~\mathrm{cm}^{-2}$ instead of $10^{20}~\mathrm{cm}^{-2}$), and the scatter of data points is a bit larger.

Finally, regarding CN, if we ignore the deficit already seen in the {\tt los} models, we reach the same conclusions : In the {\tt uniform} models, a break in the slope occurs at higher H$_2$ column densities $N(\mathrm{H}_2)\sim 6~10^{20}~\mathrm{cm}^{-2}$ (instead of $N(\mathrm{H}_2)\sim 4~10^{20}~\mathrm{cm}^{-2}$ in the {\tt los} models), the slope is definitely steeper in that high-column-density regime, and the maximum column densities reached are smaller, here by a factor $\sim 4$.

\subsection{Dark molecular gas fraction}
\label{sec:darkgas}

\begin{figure}
   \centering
   \includegraphics[width=10cm]{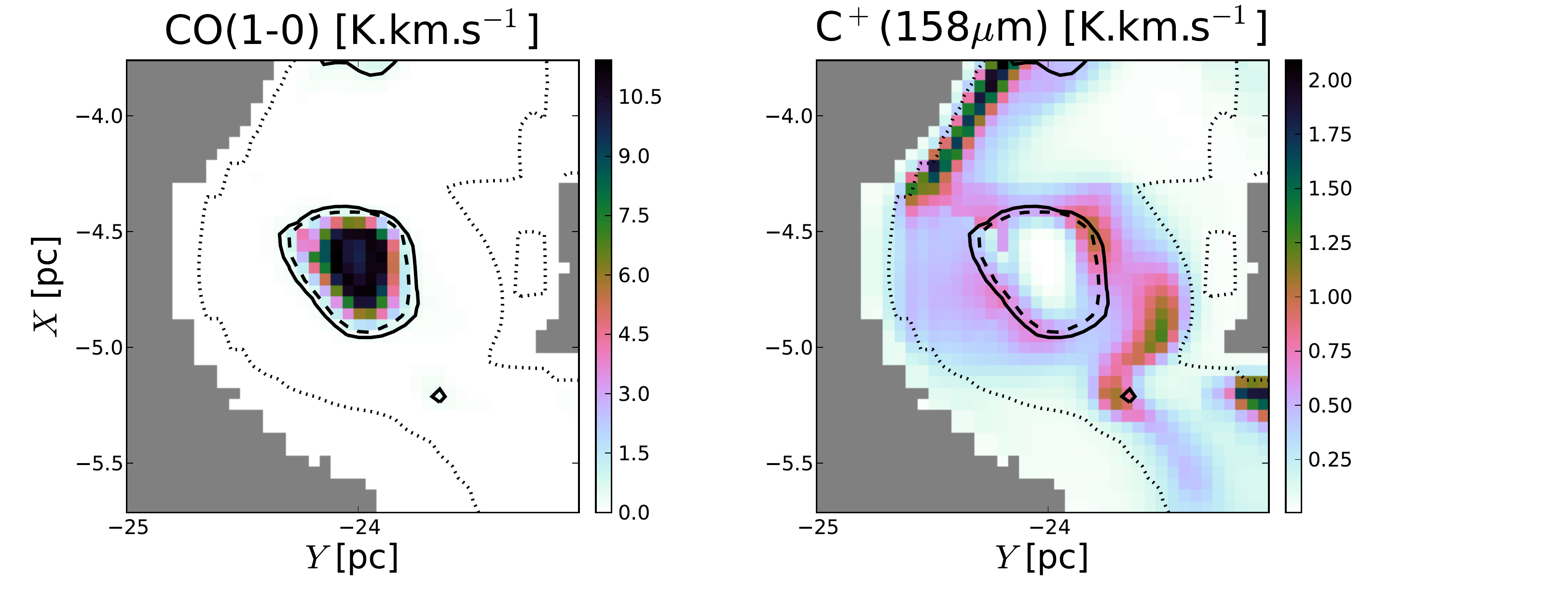}
      \caption{Synthetic emission maps in CO($J=1\to 0$) (left) and [CII] at 158 $\mu$m (right). The solid contour marks the assumed 0.4~K.km.s$^{-1}$ detection threshold for CO, the dashed contour marks the position of line center optical depth $\tau_\mathrm{CO}=1$, and the dotted contour marks the position of the molecular transition $f_\mathrm{H_2}=1/2$. Grey areas are outside of the computational domain.}
      \label{fig:COemission}
   \end{figure}

To estimate the amount of molecular gas not seen in CO - the so-called "dark gas" \citep{grenier05,planck2011} - in our simulation, we need to compute the CO line emission and compare its spatial distribution with that of H$_2$. To that effect, we focus on the cloudlet at $(X\simeq-4.7~\mathrm{pc},Y\simeq -24~\mathrm{pc})$, where the gas density peak is found, and assume a very crude cylindrical cloud model by replicating the density maps along the $Z$ axis over a line-of-sight $L=1~\mathrm{pc}$, which roughly corresponds to the cloudlet's extent in the $(X,Y)$ plane. This effectively yields column density maps which we can use with the RADEX radiative transfer code \citep{vandertak07} to obtain emission maps in the CO ($J=1\to 0$) rotational transition line at 115.271 GHz and in the [CII] fine structure transition line at 158 $\mu$m. To be precise, at each position $(X,Y)$, we treat radiative transfer along $Z$ in a plane-parallel slab geometry. RADEX works with the escape probability formalism \citep{sobolev60}, which requires specification of the line width $\Delta V$. We estimate it to be $\sigma_\mathrm{3D}/\sqrt{3}=1~\mathrm{km.s}^{-1}$, where $\sigma_\mathrm{3D}=1.8~\mathrm{km.s}^{-1}$ is the total gas velocity dispersion listed in Table~\ref{table:2}. Indeed, for any species $\alpha$ with molecular weight $\mu_\alpha$ ($\mu_{\mathrm{CO}}=28$ and $\mu_{\mathrm{C^+}}=12$), the ratio of thermal to one-dimensional turbulent velocity dispersions reads
$$
3\frac{\sigma^2_\mathrm{th}(\alpha)}{\sigma_\mathrm{3D}^2}\simeq 2.1\times\left(\frac{T}{270~\mathrm{K}}\right)\left(\frac{1}{\mu_\mathrm{A}}\right).
$$
In the cloudlet under study, $T\lesssim 100~\mathrm{K}$, so the above ratio is typically $\lesssim 0.03$ for CO and $\lesssim 0.06$ for C$^+$. It is thus reasonable to take $\Delta V=\sigma_\mathrm{3D}/\sqrt{3}$ for all RADEX runs. The code also requires specification of the gas kinetic temperature and the densities of collisional partners (H$_2$ for CO; H$_2$, H and electrons for [CII]), which we get from the PDR code outputs. RADEX is thus run on every line of sight parallel to the $Z$ axis, and results are combined into a CO ($J=1\to 0$) emission map and a [CII] emission map, both shown on Fig.~\ref{fig:COemission}.

Between the solid and dotted contours is the "dark molecular gas" region where hydrogen is predominantly in its molecular form but CO emission fails to detect it. We assume a detection threshold $W_\mathrm{CO}=0.4~\mathrm{K.km.s^{-1}}$ consistent with the noise level in e.g. the CO survey of Taurus by \citet{goldsmith08}. On the other hand, that same gas can definitely be traced in the [CII] line, as it has a typical integrated emission of $\sim 0.4-0.8~\mathrm{K.km.s^{-1}}$, while the sensitivity quoted by \citet{velusamy10} for the GOTC+ key program is $\sim 0.1-0.2~\mathrm{K.km.s^{-1}}$.

The "dark molecular gas" fraction associated with this cloudlet can be estimated by taking one-dimensional cuts parallel to the $Y$ axis going through the CO emission region. Along such a cut, which is parametrized by $X$, we define $Y_0(X)$ and $Y_1(X)$ as the boundaries of the computational domain\footnote{The dependence of $f_\mathrm{DG}$ on the boundaries of the computational domain is necessarily small, as there is little mass at low densities.} (sharp transition from grey to white on the panels of Fig.~\ref{fig:COemission}), and we note $W_\mathrm{CO}(X)$ the region where the integrated emission of CO ($J=1\to 0$) exceeds the detection threshold $W_\mathrm{CO}$ (region enclosed by the solid contour on Fig.~\ref{fig:COemission}). We then define the "dark gas" fraction as
$$
f_{\mathrm{DG}}(X)=1-\frac{\displaystyle \int_{W_\mathrm{CO}(X)} \!\!\!\!\!\!\!\!\!\! n({\mathrm{H}_2})\mathrm{d} Y}{\displaystyle \int_{Y_0(X)}^{Y_1(X)} \!\!\!\!\!\!\!\!\!\!  n({\mathrm{H}_2})\mathrm{d}Y}
$$
\begin{figure}
   \centering
   \includegraphics[width=10cm]{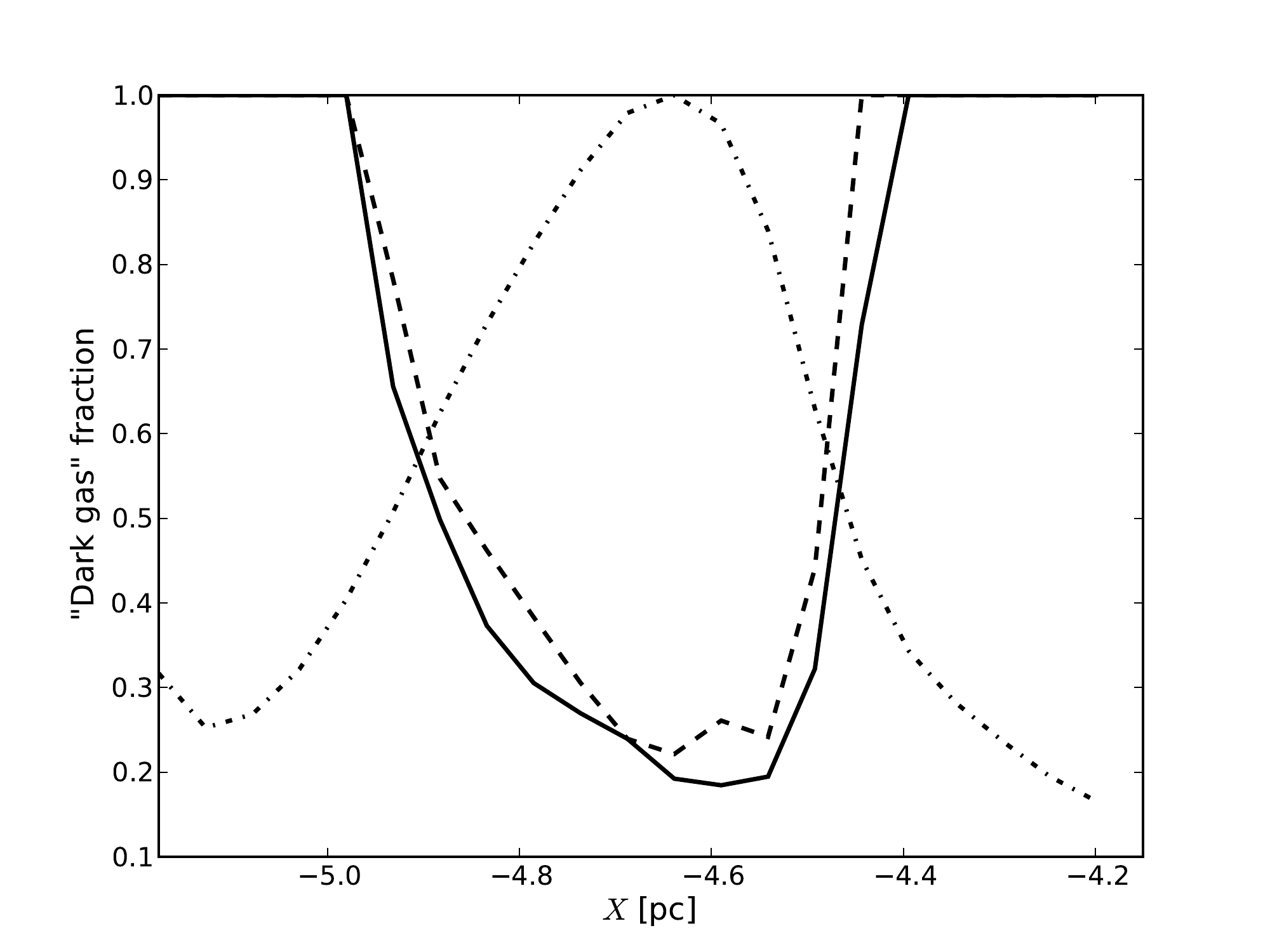}
      \caption{Fraction of "dark gas" (solid line) along one dimensional cuts parallel to the $Y$ axis going through the cloudlet at $(X\simeq-4.7~\mathrm{pc},Y\simeq -24~\mathrm{pc})$. Also shown are the fraction of "dark gas" computed using the definition by \citet{wolfire10} (dashed line), and a normalized profile of the total gas column densities $N_H$ along the same cuts (dash-dotted line).}
      \label{fig:DG}
   \end{figure}
Fig.~\ref{fig:DG} shows this fraction as a function of the position $X$ of the one-dimensional cut. Obviously, $f_\mathrm{DG}=1$ when the cut does not pass through the CO emission region, and $f_\mathrm{DG}<1$ when some of the H$_2$ is traced by CO. We find that in this cloudlet, at least 20\% of  H$_2$ is not traced by CO, even at the peak of the gas density. To get a mean fraction of dark gas in this cloudlet, we average $f_\mathrm{DG}$ over the range of $X$ coordinates where CO is seen (i.e. $f_{\mathrm{DG}}(X)<1$), weighted by the total gas column density $N_H$. This yields $\overline{f_\mathrm{DG}}=0.32$, which is somewhat higher than the findings of \citet{velusamy10}, who identified 53 "transition clouds" with both {\sc Hi} and $^{12}$CO emission but no $^{13}$CO, and found that $\sim 25\%$ of H$_2$ in these clouds belong to an H$_2$/C$^+$ layer not seen in CO. However, the scatter in observational values is large \citep{grenier05,abdo10}, so the small discrepancy is no cause for alarm. Another possible comparison is with \citet{wolfire10}, who constructed spherical models of molecular clouds to study the dark gas fraction, which they define in a similar way, except that the boundary of their CO region is specified by the condition of unit optical thickness at the line center, $\tau_\mathrm{CO}=1$. In our clump, that isocontour is very close to our own condition $I_\mathrm{CO}= W_\mathrm{CO}$, as can be seen on Fig~\ref{fig:COemission}. Computing the average dark gas fraction with their condition yields $\overline{f_\mathrm{DG}}=0.36$, which is quite close to their results $f_\mathrm{DG}\sim 0.25-0.33$. There is a notable difference between their models and ours, however, since our cloudlet has a mass $\sim 9.5~M_\odot$ inside the CO region, while \citet{wolfire10} study GMCs with masses in the range $10^5~M_\odot$ to $3~10^6~M_\odot$. Their impinging UV field is also notably higher ($\chi=3-30$).

\section{Discussion and summary}
\label{sec:conclusions}

\subsection{Illumination effects}
\begin{figure}
   \centering
   \includegraphics[width=9.2cm]{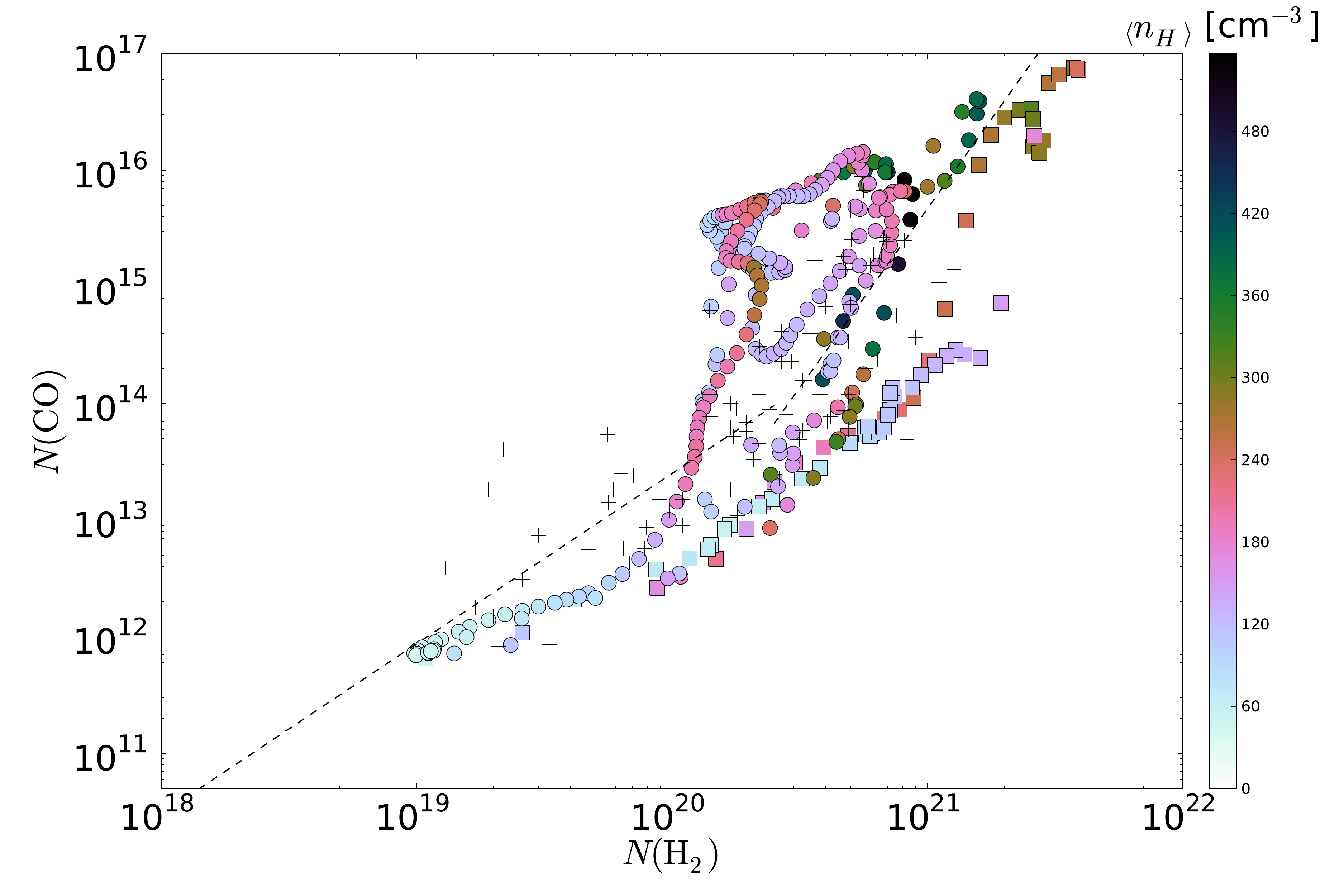}
   
      \caption{CO column densities versus H$_2$ column densities, in the {\tt los} models when combining data based on a criterion of minimum local illumination. Symbols are the same as on Fig.~\ref{fig:CO-H2} (top panel).}
      \label{fig:CO-H2-alternative}
   \end{figure}

In this paper, we bypass the one-dimensionality of the PDR code by combining runs in two orthogonal directions, taking, at each grid point, the chemical composition corresponding to maximum radiation energy density $E$. It should be noted that this choice is questionable~: if a position is shielded from radiation in almost every direction but for one small hole, the illumination at this location resulting from our procedure is too high. As this means forming less molecules, we wish to estimate if it might account for some of the CO deficit seen on Fig.~\ref{fig:CO-H2}. To do so in a simple way, we compute the chemical composition in the opposite assumption, i.e. based on the criterion of minimum local illumination. The result is plotted on Fig.\ref{fig:CO-H2-alternative}, and shows how indeed this helps recovering observed CO column densities for $N(\mathrm{H}_2)\gtrsim 10^{20}~\mathrm{cm}^{-2}$, with a consistent scatter. Below $N(\mathrm{H}_2)\simeq 10^{20}~\mathrm{cm}^{-2}$, a significant CO deficit remains, however.

Obviously, this choice of minimum local illumination is also unphysical, and the reality must lie somewhere in between. A physically better, but more computationally intensive method is being pursued and will be presented in a future paper.


\subsection{FGK approximation}

Our study uses the \citet{federman79} (FGK) approximation to compute self-shielding. This may underestimate the shielding of CO by molecular hydrogen lines, so we perform a few runs of the PDR code using exact radiative transfer \citep{goicoechea07}. We do this on some lines of sight for which $N(\mathrm{H}_2)\sim 10^{20}~\mathrm{cm}^{-2}$, to see whether this helps fill the CO deficit in that region. It turns out that the CO column densities so obtained are indeed higher than those found in the FGK approximation, but by a factor $\lesssim 2$, which is not enough to explain our CO deficit. As the computational time is on the other hand increased by a factor $\sim 5 - 6$, we feel that this approach is not to be pursued.





\subsection{Steady-state assumption}

In this study, the simulation cube is taken as a static background, under the assumption that timescales for chemistry and photoprocesses are much shorter than those of the MHD simulation.

From the analysis performed by \citet{lepetit06} on uniform density PDRs, it appears that timescales for H$_2$ photodissociation at the edges of a cloud are $\sim 1000/\chi$ yr, where $\chi$ is the FUV radiation strength in units of the \citet{draine78} field. In the analysis presented, we choose $\chi=1$ so that the corresponding timescale is about 1000 yr.

Estimating timescales for the MHD simulation is more difficult, because what we're actually interested in is the time over which structures remain coherent, and we do not have access to this information due to the Eulerian nature of the simulation, which makes it impossible to confidently identify structures. For a rough estimate, we may consider the overall crossing time $\tau_\mathrm{cross}=L/V_X\simeq 2.4~\mathrm{Myr}$, but this does not correspond to the time over which gas is mixed by turbulence at a given scale. For this, we may use the velocity dispersion $\sigma_\mathrm{3D}$ listed in Table~\ref{table:2} within the observational clump, whose size is about 2~pc. This yields a dynamical timescale $\tau_\mathrm{dyn}\simeq 1.1~\mathrm{Myr}$, which is very similar to the values quoted by \citet{wolfire10} to validate the steady-state assumption in their models.

The chemical timescale is that of the formation of molecular hydrogen. As shown by \citet{glover07} using numerical simulations of decaying ISM turbulence that include a simplified chemical network, the formation timescale for H$_2$ in turbulent magnetized molecular clouds is $\tau_\mathrm{chem}\sim 1-2~\mathrm{Myr}$. It is therefore of the order of the estimated dynamical timescales in our simulation, so that our steady-state assumption seems only marginally valid. However, H$_2$ is formed in dense regions and transported in the entire volume through turbulent motions, so we may be safe assuming steady state, provided we consider a late enough snapshot. Indeed, if H$_2$ starts forming when the converging WNM flows collide near the midplane ($\tau_\mathrm{coll}\simeq 1~\mathrm{Myr}$), and if it is fully formed and transported in the entire volume after $\tau_\mathrm{chem}+\tau_\mathrm{dyn}+\tau_\mathrm{cross}$, this requires taking a snapshot timed at no earlier than 5.4~Myr, which is the case here ($t=7.35~\mathrm{Myr}$). We conclude that our steady-state assumption is a legitimate one.

\subsection{Warm chemistry}

It should be noted that chemistry is here driven by UV radiation only, but that there is an important pathway for the formation of many molecular species, which is warm chemistry in turbulence dissipation regions (TDR), studied by \citet{joulain98} and \citet{godard09}. In particular, the CO abundances found in the models by \citet{godard09} are larger than in corresponding PDR models, sometimes by almost an order of magnitude. More generally, \citet{godard09} argue that observed chemical abundances are on the whole well reproduced if dissipation is due to ion-neutral friction in sheared structures $\sim 100~\mathrm{AU}$ thick. A TDR post-processing of our MHD simulation is in the works, to compare both types of chemistry.

\subsection{Summary}

We have presented a first analysis of UV-driven chemistry in a simulation of the diffuse ISM, by post-processing it with the Meudon PDR code. Our results show that assuming a uniform density medium when modelling PDRs leads to significant errors~: in the case of CO, for instance, the maximum column densities found with this simplistic assumption are a factor $\sim 3$ lower than those found using actual density fluctuations. The slope of the H$_2$-CO correlation at $N(\mathrm{H}_2)\gtrsim 5~10^{20}~\mathrm{cm}^{-2}$ is also a factor $\sim 3$ higher than in the more realistic case, and therefore much less in agreement with observations. A second result of our study is that, in the densest parts of the simulation ($n_H\gtrsim 10^3~\mathrm{cm}^{-3}$), some 35\% to 40\% of the molecular gas is "dark", in the sense that it it not traced by the CO($J=1\to 0$) line, given current sensitivities. It is however detectable via the [CII] fine structure transition line at 158~$\mu$m. As a side result, we find that the simplified cooling used in the MHD simulation by~\citet{hennebelle08} yields gas temperatures in reasonable agreement with those found using the more detailed processes included in the PDR code.

\begin{acknowledgements}
The authors acknowledge support for computing resources and services from France Grilles and the EGI e-infrastructure. Some kinetic data have been downloaded from the online KIDA (KInetic Database for Astrochemistry, {\tt http://kida.obs.u-bordeaux1.fr}) database. Colour figures in this paper use the {\tt cubehelix} colour map by~\citet{green2011}.
\end{acknowledgements}

\appendix

\section{Using the Meudon PDR code with density profiles}
\label{sec:PDR-wdp}

This appendix is meant as an introduction to using the Meudon PDR code\footnote{We use version 1.4.1 of the PDR code, with a fixed H$_2$ formation rate $R_f=3~10^{-17}\sqrt{T/100~\mathrm{K}}$.} with fluctuating density profiles. For more detailed presentations of the code, the reader is referred to \citet{lebourlot99,lepetit06,gonzalez08}. 

The code requires two input files : a {\tt .pfl} file listing visual extinction $A_V$, temperature $T$ (in K) and total gas density $n_H$ (in $\mathrm{cm}^{-3}$) along the line of sight, and a {\tt .in} file supplying the parameters of the run to perform. These are listed in Table~\ref{table:1}, and some of them require a short comment~:
\\
- {\tt modele} is the basename chosen for output files. For the {\tt los} models, we use a name of the generic form {\tt los\_z$Z$\_y$Y$\_x$X_m$-$X_p$} or {\tt los\_z$Z$\_x$X$\_y$Y_m$-$Y_p$}, reflecting the position of the extracted profile in the cube. Note that $Z$ is a constant throughout this paper, as is evident from Fig.~\ref{FigColDensClump}. For the {\tt uniform} models, we use names of the generic form {\tt uniform\_z$Z$\_y$Y$\_x$X_m$-$X_p$} or {\tt uniform\_z$Z$\_x$X$\_y$Y_m$-$Y_p$}.
\\
- {\tt ifafm} is the number of global iterations to use. As \citet{lepetit06} point out, for diffuse clouds ($A_V < 0.5$) proper convergence may require up to 20 iterations, so we select {\tt ifafm=20} for all of our models. For information, 415 of the 447 profiles have total $A_V < 0.5$.
\\
- {\tt Avmax} is that same total visual extinction through the cloud, which is simply the last $A_V$ value in the {\tt .pfl} file.
\\
- We set the density {\tt densh}, temperature {\tt tgaz}, and pressure {\tt presse=densh$\times$tgaz} parameters to the average values\footnote{i.e. density-weighted averages for temperature and pressure.} for each profile. They are not used by the code with a density-temperature profile, but they are used, in the reference {\tt uniform} models, as initial guesses for thermal balance computation.
\\
- {\tt radm} and {\tt radp} specify the strengths $\chi_m$ and $\chi_p$ of the incident radiation field in units of the ISRF, respectively on the left and right sides of the profile. For the runs described in this paper, we use $\chi_m=\chi_p=1$. 
\\
- {\tt fprofil} specifies the {\tt .pfl} density-temperature profile file. For consistency, we use the same naming scheme as for {\tt modele}.
\\
- {\tt vturb} is the "turbulent velocity dispersion". It does not include thermal dispersion, so we take it to be the standard deviation of the line-of-sight velocity within each structure, noted $\sigma_X$ and $\sigma_Y$ in Tables~\ref{table:0} and~\ref{table:0bis}, respectively.
\\
-{\tt ifisob} is a flag specifying whether to use a density profile. For the {\tt los} models, we therefore set {\tt ifisob=1}, to enforce the use of a density-temperature {\tt .pfl} file. However, since thermal balance is solved ({\tt ieqth=1}), the temperature values in the file are only used as initial guesses. For the {\tt uniform} models, we set {\tt ifisob=0} to use a constant density (specified by {\tt densh}).




\begin{table}
\caption{Parameters used in the PDR code, as input in the {\tt .in} files.}
\label{table:1}
\centering
\begin{tabular}{l l l}
\hline\hline          
Parameter & Description  &  Value\\
\hline
{\tt modele} & Basename for the output files & {\it see appendix~\ref{sec:PDR-wdp}}  \\
{\tt chimie} & Chemistry file & {\tt \tiny chimie08}~$^{\mathrm{a}}$ \\
{\tt ifafm} & Number of global iterations & 20  \\
{\tt Avmax} & Integration limit in $A_V$ & {\it see appendix~\ref{sec:PDR-wdp}}  \\
{\tt densh} & Initial density ($\mathrm{cm}^{-3}$) & {\it see appendix~\ref{sec:PDR-wdp}}  \\
{\tt F\_ISRF} & ISRF expression flag & 1~$^{\mathrm{b}}$ \\
{\tt radm} & ISRF scaling factor $\chi_m$ & {\it see appendix~\ref{sec:PDR-wdp}}  \\
{\tt radp} & ISRF scaling factor $\chi_p$ & {\it see appendix~\ref{sec:PDR-wdp}}  \\
{\tt srcpp} & Additional radiation field source & {\tt \tiny none.txt} \\
{\tt d\_sour} & Star distance (pc) & 0~$^{\mathrm{c}}$ \\
{\tt fmrc} & Cosmic rays ionisation rate ($10^{-17}~\mathrm{s}^{-1}$) & 5 \\
{\tt ieqth} & Thermal balance computation flag & 1~$^{\mathrm{d}}$ \\
{\tt tgaz} & Initial temperature (K) & {\it see appendix~\ref{sec:PDR-wdp}} \\
{\tt ifisob} & State equation flag & {\it see appendix~\ref{sec:PDR-wdp}} \\
{\tt fprofil} & Density-Temperature profile file & {\it see appendix~\ref{sec:PDR-wdp}} \\
{\tt presse} & Initial pressure ($\mathrm{K.cm}^{-3}$)& {\it see appendix~\ref{sec:PDR-wdp}}  \\
{\tt vturb} & Turbulent velocity ($\mathrm{km.s}^{-1}$) & {\it see appendix~\ref{sec:PDR-wdp}}  \\
{\tt itrfer} & UV transfer method flag & 0~$^{\mathrm{e}}$ \\
{\tt jfgkh2} & Minimum $J$ level for FGK approximation & 0 \\
{\tt ichh2} & H + H$_2$ collision rate model flag & 2~$^{\mathrm{f}}$ \\
{\tt los\_ext} & Line of sight extinction curve & {\tt \tiny Galaxy}~$^{\mathrm{g}}$ \\
{\tt rrr} & Reddening coefficient $R_V$ = $A_V / E_{B-V}$ & 3.1~$^{\mathrm{h}}$ \\
{\tt cdunit} & Gas-to-dust ratio $C_D=N_H / E_{B-V}$ ($\mathrm{cm}^{-2}$) & $5.8\times 10^{21}$~$^{\mathrm{g}}$ \\
{\tt alb} & Dust albedo & 0.42~$^{\mathrm{g}}$ \\
{\tt gg} & Diffusion anisotropy factor $\left<\cos\theta\right>$ & 0.6~$^{\mathrm{g}}$ \\
{\tt gratio} & Mass ratio of grains / gas &  0.01~$^{\mathrm{g}}$\\
{\tt rhogr} & Grains mass density ($\mathrm{g.cm}^{-3}$) & 2.59~$^{\mathrm{i}}$ \\
{\tt alpgr} & Grains distribution index & 3.5~$^{\mathrm{g}}$ \\
{\tt rgrmin} & Grains minimum radius (cm) & $3\times 10^{-7}$~$^{\mathrm{g}}$ \\
{\tt rgrmax} & Grains maximum radius (cm) & $3\times 10^{-5}$~$^{\mathrm{g}}$ \\
{\tt F\_DUSTEM} & DUSTEM activation flag & 0~$^{\mathrm{j}}$ \\
{\tt iforh2} & H$_2$ formation on grains model flag & 0~$^{\mathrm{k}}$ \\
{\tt istic} & H sticking on grain model flag & 4~$^{\mathrm{l}}$ \\
\hline    
\end{tabular}
\begin{list}{}{}
\item[$^{\mathrm{a}}$] This chemistry file does not include deuterated species.
\item[$^{\mathrm{b}}$] Uses expressions based on \citet{mathis83} and \citet{black94} rather than \citet{draine78}.
\item[$^{\mathrm{c}}$] This means that no additional star is present.
\item[$^{\mathrm{d}}$] Thermal balance is computed for each point in the cloud, using the temperatures given by the {\tt .pfl} file as initial guesses.
\item[$^{\mathrm{e}}$] Use the \citet{federman79} (FGK) approximation for the H$_2$ lines in the UV.
\item[$^{\mathrm{f}}$] Use the values compiled by \citet{lebourlot99} with reactive collisions from \citet{schofield67}.
\item[$^{\mathrm{g}}$] See Table 4 of \citet{lepetit06}, which quotes values from \citet{fitzpatrick90} for the extinction curve, \citet{bohlin78,rachford02} for $C_D$, \citet{mathis96} for the dust albedo, \citet{weingartner01} for $\left<\cos\theta\right>$, \citet{mathis77} for the grain size distribution parameters.
\item[$^{\mathrm{h}}$] See for instance \citet{cardelli89}
\item[$^{\mathrm{i}}$] Taken from \citet{gonzalez08}.
\item[$^{\mathrm{j}}$] Do not couple to the DUSTEM code \citep{compiegne10}. 
\item[$^{\mathrm{k}}$] Energy released by H$_2$ formation on dust grains is equally split between grain excitation, H$_2$ kinetic energy and internal energy. See section 6.1.2 in~\citet{lepetit06} for details.
\item[$^{\mathrm{l}}$] See Appendix E5 in~\citet{lepetit06}.
\end{list}
\end{table}






\begin{figure}
   \centering
   \includegraphics[width=8cm]{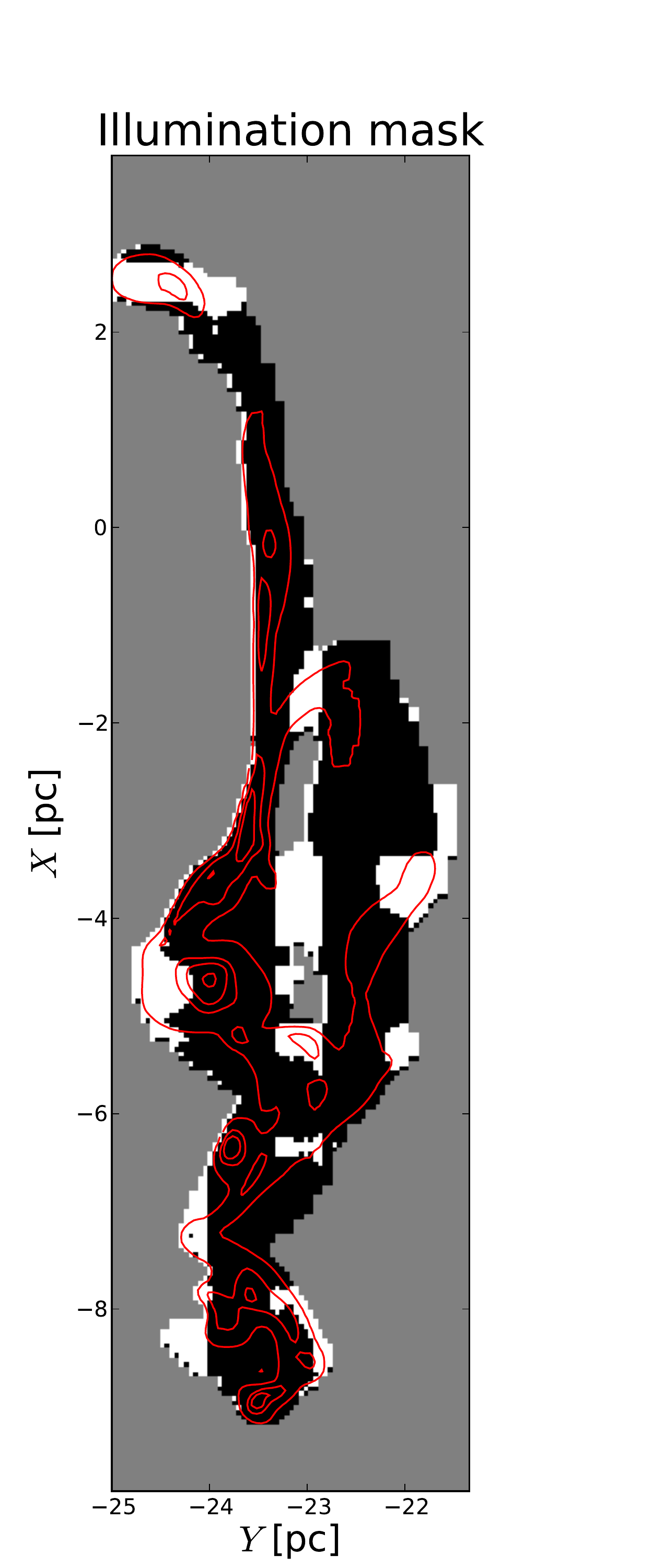}
      \caption{Illumination mask computed by comparing at each point ($X,Y$) the local radiative energy densities $E_X$ and $E_Y$ output by the PDR code along the $X$ and $Y$ directions, respectively. Regions where $E_Y\geqslant E_X$ are marked in black and regions where $E_X> E_Y$ are marked in white. Contour lines of equal total gas density are overlaid at 20, 100, 500, 1000 and 2000 cm$^{-3}$. Grey areas are outside of the computational domain.}
      \label{fig:Illumination_mask}
   \end{figure}

\section{Post-processing of raw outputs}
\label{sec:postprocessing}

\subsection{Resampling}
Outputs of the PDR code are FITS data files and XML description files, and we use dedicated scripts to extract specific quantities into plain text files for subsequent analysis. Among the quantities retrieved are the distance $d$ from the surface of the structure, visual extinction $A_V$, proton column density $N_H$, temperature $T_\mathrm{PDR}$, proton density $n_H$, pressure $p$, ionization fraction $x_e$, and abundances $n(\alpha)$ of 99 chemical species.

As the PDR code does its own mesh refinement to better solve for the H/H$_2$ transition, these quantities are sampled irregularly. Consequently, we resample outputs on the same regular grid as the MHD simulation, using a simple linear interpolation method. This allows us to build raw maps for all output quantities from PDR code runs along the $X$ and $Y$ directions.

\subsection{Missing data}
For reasons that are unclear, a few\footnote{Namely, for the {\tt los} models, 16 out of 291 along $Y$ and 3 out of 156 along $X$; for the {\tt uniform} models, 8 out of 291 along $Y$ and 6 out of 156 along $X$.} of the 894 PDR runs do not complete successfully on the EGEE grid. To supplement the missing data, we interpolate along the perpendicular direction. Consider the ensemble of runs along the $X$ direction : completed runs yield quantities $F_X(X,Y)$, and if a run is missing at coordinate $Y=Y_0$, we supply $F_X(X,Y_0)$ by linearly interpolating $G(Y)=F_X(X_0,Y)$ at constant $X=X_0$. This yields a satisfactory completion of the raw data.

   
\subsection{Combination of $X$ and $Y$ runs}
We then proceed to the combination of data from runs along the $X$ and $Y$ directions, as described in~\ref{sec:method}. Fig.~\ref{fig:Illumination_mask} shows the "illumination mask" computed by comparing the local radiation energy densities $E_X$ and $E_Y$ output by the PDR code in {\tt los} models parallel to the $X$ and $Y$  directions, respectively. This mask is then used to build a single data array for each quantity $F$ output by the PDR code, at each grid point $(X,Y)$, according to the rule :
\begin{displaymath}
F = \left\{ \begin{array}{ll}
F_X & \quad\textrm{if}\quad E_X> E_Y\\
F_Y & \quad\textrm{if}\quad E_X\leqslant E_Y
\end{array} \right.
\end{displaymath}
This helps to reduce the shadowing artifacts due to the one-dimensionality of the PDR code, while ensuring element conservation in each grid cell, and yields the final maps that are analyzed and discussed in the main body of the paper. In the discussion (section~\ref{sec:conclusions}), we also make use of the inverse choice :
\begin{displaymath}
F = \left\{ \begin{array}{ll}
F_Y & \quad\textrm{if}\quad E_X> E_Y\\
F_X & \quad\textrm{if}\quad E_X\leqslant E_Y
\end{array} \right.
\end{displaymath}

\end{document}